\documentclass[journal, final]{IEEEtran}

\usepackage{booktabs}
\usepackage{outlines}
\usepackage{xspace}
\usepackage{graphicx}
\usepackage{lipsum}
\usepackage[T1]{fontenc}
\usepackage[utf8]{inputenc}
\usepackage{times}

\usepackage{pgfplots}
    \usepgfplotslibrary{colorbrewer}
    \pgfplotsset{
        cycle list/Set1-9,
        cycle multiindex* list={
            mark list*\nextlist
            Set1-9\nextlist
        },
    }
    \usepgfplotslibrary{fillbetween}
    \usetikzlibrary{arrows}
    \usetikzlibrary{calc}
    \usetikzlibrary{patterns}
    \usetikzlibrary{calc}

\usepackage{multirow}
\usepackage{subcaption}
\usepackage{amsmath}
\usepackage{amsfonts}
\usepackage{amssymb}
\usepackage{algorithm}
\usepackage{algpseudocode}
\usepackage[acronym]{glossaries}
\usepackage[numbers]{natbib}
\usepackage{dblfloatfix}
\usepackage{hyperref}
\hypersetup{
    colorlinks=true,
    linkcolor=blue,
    filecolor=magenta,      
    urlcolor=cyan,
}
\usepackage{cleveref}
\crefname{section}{\S}{sections}
\Crefname{section}{\S}{Sections}
\crefname{figure}{Fig.}{Figs.}
\Crefname{figure}{Fig.}{Figs.}
\crefname{equation}{}{equations}
\Crefname{equation}{Eq.}{Equations}
\newcommand{\argmin}{\arg\!\min}
\newcommand{\opt}{Maximum-Likelihood\xspace}
\newcommand{\sys}{MMNet\xspace}
\newcommand{\sysiid}{MMNet-iid\xspace}
\newcommand{\detnet}{DetNet\xspace}
\newcommand{\xh}{\hat{\textbf{x}}}
\newcommand{\x}{\textbf{x}}
\newcommand{\xb}{\textbf{x}}
\newcommand{\y}{\textbf{y}}
\newcommand{\z}{\textbf{z}}
\newcommand{\yb}{\textbf{y}}

\newcommand{\n}{\textbf{n}}
\newcommand{\Hb}{\textbf{H}}
\newcommand{\elin}{\textbf{e}^{lin}}
\newcommand{\eden}{\textbf{e}^{den}}
\newcommand{\real}{3GPP MIMO\xspace}

\newacronym{iid}{i.i.d.}{independent and identically distributed}
\newacronym{amp}{AMP}{approximate message passing}
\newacronym{ist}{IST}{iterative soft-thresholding}
\newacronym{oamp}{OAMP}{orthogonal AMP}
\newacronym{ml}{ML}{maximum likelihood}
\newacronym{snr}{SNR}{signal-to-noise ratio}
\newacronym{qam}{QAM}{quadrature amplitude modulation}
\newacronym{bpsk}{BPSK}{binary phase shift keying}
\newacronym{sdr}{SDR}{semidefinite relaxation}
\newacronym{zf}{ZF}{zero-forcing}
\newacronym{mmse}{MMSE}{minimum mean square error}
\newacronym{bp}{BP}{belief propagation}
\newacronym{bs}{BS}{base station}
\newacronym{ser}{SER}{symbol error rate}
\newacronym{fp32}{FP32}{32bits floating-point}

\title{Adaptive Neural Signal Detection\\ for Massive MIMO}

\author{
Mehrdad Khani,
Mohammad Alizadeh,
Jakob Hoydis, \IEEEmembership{Senior Member, IEEE},
Phil Fleming, \IEEEmembership{Senior Member, IEEE}
\thanks{M. Khani and M. Alizadeh are with the Computer Science \& Artificial Intelligence Laboratory, MIT, Cambridge, MA 02139, USA (khani@mit.edu, alizadeh@csail.mit.edu).
}
\thanks{J. Hoydis is with Nokia Bell Labs, Paris-Saclay, 91620 Nozay, France (jakob.hoydis@nokia-bell-labs.com).
}
\thanks{P. Fleming is an independent technology consultant (pfleming91@gmail.com).}
}

\begin{document}
    \maketitle
    \begin{abstract}
Symbol detection for Massive Multiple-Input Multiple-Output (MIMO) is a challenging problem for which traditional algorithms are either impractical or suffer from performance limitations. Several recently proposed learning-based approaches achieve promising results on simple channel models (e.g., i.i.d. Gaussian). However, their performance degrades significantly on real-world channels with spatial correlation. We propose \sys, a deep learning MIMO detection scheme that significantly outperforms existing approaches on realistic channels  with the same or lower computational complexity. \sys's design builds on the theory of iterative soft-thresholding algorithms and uses a novel training algorithm that leverages temporal and spectral correlation to accelerate training. Together, these innovations allow \sys to train online for every realization of the channel. On i.i.d. Gaussian channels, \sys requires two orders of magnitude fewer operations than existing deep learning schemes but achieves near-optimal performance. On spatially-correlated channels, it achieves the same error rate as the next-best learning scheme (OAMPNet~\cite{he2018model}) at 2.5dB lower \glsfirst{snr} and with at least 10$\times$ less computational complexity. \sys is also 4--8dB better overall than a classic linear scheme like the minimum mean square error (MMSE) detector.


\end{abstract}

    \section{Introduction}
The fifth generation of cellular communication systems (5G) promises an order of magnitude higher spectral efficiency (measured in bits/s/Hz) than legacy standards such as Long Term Evolution (LTE)~\cite{osseiran20165g}. One of the key enablers of this better efficiency is Massive Multiple-Input Multiple-Output~(MIMO)~\cite{massivemimobook}, in which a \gls{bs} equipped with a very large number of antennas (around 64--256) simultaneously serves multiple single-antenna user equipments (UEs) on the same time-frequency resource. 

Legacy systems already use MIMO~\cite{telatar1999capacity}, but this is the first time it will be deployed on such a large scale, creating significant challenges for {\em signal detection}.  The goal of signal detection is to infer the transmitted signal vector $\bold{x}$ from the vector $\bold{y} = \bold{H}\bold{x} + \bold{n}$ received at the BS antennas, where $\bold{H}$ is the channel matrix and $\bold{n}$ is Gaussian noise. Traditional MIMO detection methods with strong performance~\cite{viterbo1999universal, yang2015fifty, larsson2009mimo, wiesel2005semidefinite} are feasible only for small systems and have prohibitive complexity for massive MIMO deployments. Thus, there is a  need for low-complexity symbol detection schemes that perform well and scale to large system dimensions.

In recent work, researchers have proposed several learning approaches to MIMO detection. \citet{samuel2017deep} developed a deep neural network architecture called \detnet with impressive performance, e.g., matching the performance of a \gls{sdr} baseline for \gls{iid} Gaussian channel matrices 
while running 30$\times$ faster. Shortly afterwards, inspired by the Orthogonal AMP algorithm~\cite{ma2017orthogonal}, \citet{he2018model} introduced OAMPNet and demonstrated strong performance on both \gls{iid} Gaussian and small-sized correlated channel matrices based on the Kronecker model with exponentially-distributed spatial correlations~\cite{loyka2001channel}. \detnet and OAMPNet are both trained offline: they try to learn a single model during training for a family of channel matrices (e.g., \gls{iid} Gaussian channels). However, the two schemes have different design philosophies. \detnet embeds little domain knowledge into the model and relies on a large neural network with 1-10 million parameters depending on the system size and modulation scheme. By contrast, OAMPNet takes a model-driven approach and follows the OAMP algorithm closely; it adds only 2 trainable parameters per iteration of the OAMP algorithm. 

In this paper we show that neither approach is effective in practice. We conduct extensive experiments using a dataset of channel realizations from the 3GPP 3D MIMO channel~\cite{3gpp-3d-mimo}, as implemented in the QuaDRiGa channel simulator~\cite{jaeckel2014quadriga}. Our results show that DetNet's training is unstable for realistic channels, while OAMPNet suffers a large performance gap (5--7dB at symbol error rate of $10^{-3}$) compared to the optimal Maximum-Likelihood detector on these channels.
Both models (as well as several classical baselines) perform well in simpler settings used for evaluation in prior work (e.g., \gls{iid} Gaussian channels, low-order modulation schemes). Our results demonstrate the difficulty of learning a fixed detector that generalizes across a wide variety of channel matrices (esp. poorly-conditioned channels that are difficult to invert). \detnet's approach is, in a sense, {\em too} general, making the large model difficult to train, while OAMPNet makes strong assumptions about channel matrices (OAMP was designed for unitarily-invariant channels~\cite{ma2017orthogonal}) and, therefore, performs poorly on channels that deviate from these assumptions.

Motivated by these findings, we revisit MIMO detection from an online learning perspective. We ask: {\em Can a receiver adapt its detector for every realization of the channel matrix?} Such an approach would arguably be simpler and could perform better than a fixed detector that must handle a wide variety of channel matrices. However, conventional wisdom suggests that training a deep neural network online is ``impossible'' in this context because of the stringent performance requirements of MIMO detectors~\cite{samuel2017deep}.

\sys overcomes this challenge with two key ideas. First, it uses a neural network architecture that strikes a balance between expressivity and complexity. \sys's neural network is based on iterative soft-thresholding algorithms~\cite{beck2009fast, jeon2015optimality}. It preserves important aspects of these algorithms in MIMO detection, such as a denoiser architecture tailored for uncorrelated Gaussian noise for different transmitted signals. At the same time, \sys introduces adequate flexibility into these algorithms, with trainable parameters that are optimized for each channel realization. Second, \sys's online training algorithm exploits the locality of channel matrices at a receiver in both the frequency and time domains.  By leveraging locality, \sys accelerates training 250$\times$ compared to naively retraining the neural network from scratch for each channel realization. Taken together, these ideas enable \sys to achieve performance within 1.5dB of the optimal Maximum-Likelihood detector with 10-15$\times$ less computational complexity than the second best scheme, OAMPNet. On random \gls{iid} Gaussian channels, we show that a simple version of \sys with 100$\times$ less complexity than OAMPNet and \detnet, achieves near-optimal performance without requiring any retraining.   

We empirically analyze the dynamics of errors across different layers of \sys and OAMPNet to understand why \sys achieves higher detection accuracy. Our analysis reveals that \sys shapes the distribution of noise at the input of denoisers to ensure they operate effectively. In particular, as signals propagate through the \sys neural network, the noise distribution at the input of denoiser stages approaches a Gaussian distribution, to create precisely the conditions in which the denoisers can attenuate noise maximally.

The rest of this paper is organized as follows. \Cref{sec:background} provides background on classical and learning-based detection schemes, and introduces a general iterative framework that can express many of these algorithms. \Cref{sec:design} introduces the \sys design in addition to a simple variant for \gls{iid} channels. \Cref{sec:experiments} shows performance results of detection algorithms on \gls{iid} Gaussian and \real channels for different modulations. \Cref{sec:error_analysis} discusses the error dynamics of \sys and empirically studies why it performs better than OAMPNet. \Cref{sec:cost} introduces the \sys online training algorithm and how temporal and spectral locality of channel matrices can significantly reduce the computational complexity of training \sys. We release our Tensorflow~\cite{tensorflow2015-whitepaper} implementation of learning-based schemes, spatially correlated channels dataset, and benchmark schemes at \href{https://github.com/mehrdadkhani/MMNet}{https://github.com/mehrdadkhani/MMNet}. 

\noindent {\bf Notation:} We will use lowercase symbols for scalars, bold lowercase symbols for column vectors and bold uppercase symbols to denote matrices. Symbols $\{\theta, \boldsymbol{\theta}, \boldsymbol{\Theta}\}$ are used to represent the parameters of trainable models. The pseudo-inverse of the matrix $\textbf{H}$ is denoted by $\Hb^+=(\Hb^H\Hb)^{-1}\Hb^H$. $\textbf{I}_n$ stands for identity matrix of size $n$.

    \section{Background and Related Work}
\label{sec:background}
This section introduces the MIMO detection problem and reviews the most relevant related work.

\subsection{Problem Definition}
We consider a communication channel from $N_t$ single-antenna transmitters to a receiver equipped with $N_r$ antennas. The received vector $\textbf{y}\in\mathbb{C}^{N_r}$ is given as
\begin{align}\label{eq:channel}
    \y = \Hb \x + \n
\end{align}
where $\textbf{H}\in \mathbb{C}^{N_r\times N_t}$ is the channel matrix, $\textbf{n}\sim\mathcal{C}\mathcal{N}(0,\sigma^2\textbf{I}_{N_r})$ is complex Gaussian noise, and $\textbf{x} \in\mathcal{X}^{N_t}$ is the vector of transmitted symbols.  $\mathcal{X}$ denotes the finite set of constellation points. We assume that each transmitter chooses a symbol from $\mathcal{X}$ uniformly at random, and all transmitters use the same constellation set. Further, as is standard practice, we assume that the constellation set $\mathcal{X}$ is given by a \glsfirst{qam} scheme~\cite{hanzo2004quadrature}. All constellations are normalized to unit average power (e.g., the QAM4 constellation is $\{\pm\frac{1}{\sqrt{2}}\pm j\frac{1}{\sqrt{2}} \}$).

The channel matrix $\textbf{H}$ is generated by a stochastic process, but it is assumed to be perfectly known at the receiver. The goal of the receiver is to compute the \glsfirst{ml} estimate $\hat{\textbf{x}}$ of $\textbf{x}$:
\begin{equation}\label{eq:problem}
    \hat{\textbf{x}}= \argmin_{\textbf{x}\in \mathcal{X}^{N_t}} ||\textbf{y}-\textbf{Hx}||_2.
\end{equation}

The optimization problem in \cref{eq:problem} is NP-hard due to the finite-alphabet constraint $\textbf{x}\in \mathcal{X}^{N_t}$~\cite{del2017mixed}. Over the last three decades, researchers have proposed a variety of detectors for this problem with differing levels of complexity.
We briefly describe a small subset of existing detection schemes in this section. We refer the interested reader to \cite{yang2015fifty, larsson2009mimo} for a comprehensive overview of MIMO detection schemes.

\subsection{An iterative framework for MIMO detection}
\label{sec:iter-framework}
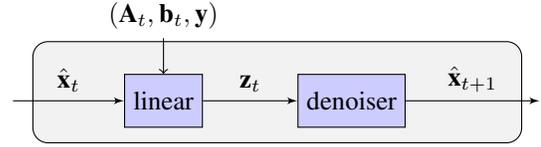
\begin{figure}
    \centering
    \usetikzlibrary{arrows}
\usetikzlibrary{fit}					
\usetikzlibrary{backgrounds}	

\tikzstyle{int}=[draw, fill=blue!20, minimum size=2em]
\tikzstyle{init} = [pin edge={to-,thin,black}]
\tikzstyle{background}=[rectangle, 
                                                fill=gray!10,
                                                inner sep=0.2cm,
                                                rounded corners=2mm, draw=black]
\begin{tikzpicture}[node distance=2.5cm,auto,>=latex']
    \node [int, pin={[init]above:$(\textbf{A}_t,\textbf{b}_t, \yb)$}] (a) {linear};
    \node (b) [left of=a,node distance=2cm, coordinate] {a};
    \node [int] (c) [right of=a] {denoiser};

    \node [coordinate] (end) [right of=c, node distance=2.5cm]{};
    \path[->] (b) edge node (input) {$\xh_t$} (a);
    \path[->] (a) edge node {$\textbf{z}_t$} (c);
    \draw[->] (c) edge node (output){$\xh_{t+1}$} (end) ;
\begin{pgfonlayer}{background}
\node [background,
                    fit=(input) (output) (a)] {};
\end{pgfonlayer}

\end{tikzpicture}
    \caption{A block of an iterative detector in our general framework. Each block contains a linear transformation followed by a denoising stage.}
    \label{fig:my_label}
\end{figure}

We focus on a class of iterative estimation algorithms for solving \cref{eq:problem} as shown in \cref{fig:my_label}. Each iteration of these algorithms comprises the following two steps:
\begin{align}\label{eq:general_fmwk}
\text{General Iteration:}&&
\begin{split}
    \textbf{z}_t &= \xh_{t} + \textbf{A}_t (\y - \Hb\xh_t) +\textbf{b}_t\\
    \xh_{t+1} &=\eta_t\left( \textbf{z}_t \right).
\end{split}
\end{align}
The first step takes as input $\xh_t$, a current estimate of $\x$ and the received signal $\y$ and applies a {\em linear} transformation to obtain an intermediate signal $\z_t$. In the second step, a {\em non-linear} ``denoiser'' is applied to $\z_t$ to produce $\xh_{t+1}$, a new estimate of $\x$ that is used as the input for the next iteration. Together, the linear and denoising operations aim to improve the quality of the estimate $\xh_t$ from one iteration to the next. 

We refer to $\y - \Hb\xh_t$ as the \emph{residual} term.
The denoiser $\eta_t(\cdot)$ can be any non-linear function in general; however, most algorithms apply the same {\em thresholding} function $\beta_t \colon \mathbb{C} \to \mathbb{C}$ to each element. Using an element-wise thresholding function can significantly reduce the complexity of the denoising step. Typically denoisers also require one or more scalar parameters which depend on the detector information of the system (channel measurement, residuals, etc.) and which need to be updated for each iteration of the algorithm. We denote them by $\sigma_t$. We use the terms \emph{step}, \emph{layer}, and \emph{block} interchangeably to refer to one complete iteration (the linear step followed by the non-linear denoiser) of the algorithms. All algorithms discussed assume $\xh_0=0$.

A natural choice for the denoising function is the minimizer of $\mathbb{E}[\|\xh - \x\|_2 |\textbf{z}_t]$, which is given by:
\begin{equation}\label{eq:gen_eta}
     \eta_t(\z_t)=\mathbb{E}[\x|\textbf{z}_t].
\end{equation}

\paragraph*{Optimal denoiser for Gaussian noise}\label{ex:gaussian_denoiser} Assume that the noise at the input of the denoiser $\z_t - \x$ has an \gls{iid} Gaussian distribution with diagonal covariance matrix $\sigma_t^2 \textbf{I}_{N_t}$. The element-wise thresholding function derived from \cref{eq:gen_eta} has the form
\begin{equation}
    \beta^g_t(z;\sigma_t^2) = \frac{1}{Z}\sum_{x_i \in \mathcal{X}} x_i\exp\left(-\frac{\|z-x_i\|^2}{\sigma_t^2}\right)
    \label{eq:gaussian-denoiser}
\end{equation}
where $Z=\sum_{x_j\in\mathcal{X}}\exp\left(-\frac{\|z-x_j\|^2}{\sigma_t^2}\right)$. As we see here, $\sigma_t$ in \cref{eq:gaussian-denoiser}  represents the standard deviation of the Gaussian noise on the denoiser inputs. In all denoisers in $\eta_t(\cdot;\sigma_t^2)$ format in this paper, $\sigma_t^2$ refers to the variance of noise in denoiser input. 

In the following, we briefly describe several algorithms for MIMO detection. We begin with traditional, non-learning approaches (\cref{sec:mimo-classical}) and then discuss recent deep learning proposals (\cref{sec:mimo-learning}). We show how many of these algorithms can be expressed in the iterative framework discussed above.

\subsection{Classical MIMO detection algorithms}
\label{sec:mimo-classical}

\subsubsection{Linear}\label{sec:linear}
The simplest method to approximately solve \cref{eq:problem} is to relax the constraint of $\textbf{x}\in\mathcal{X}^{N_t}$ to $\textbf{x}\in\mathbb{C}^{N_t}$ and then round the relaxed solution to a point on the constellation: 
\begin{align}
\text{Linear:} &&
\begin{split}
        \textbf{z} &= \argmin_{\textbf{x}\in \mathbb{C}^{N_t}} \|\textbf{y}-\textbf{Hx}\|_2
    = \Hb^+ \textbf{y}\\
    \hat{\x} &= \argmin_{\x \in \mathcal{X}^{N_t}}\|\x-\textbf{z}\|_2.
\end{split}
\label{eq:linear}
\end{align}  
Rounding each component of $\textbf{z}$ to the closest point in the constellation set $\xh$ leads to the well-known \gls{zf} detector, which is equivalent to a single step of \cref{eq:general_fmwk} with initial condition of $\xh_0=0$, $\bold{A}_0=\bold{H}^+$, $\textbf{b}_0 = 0$, and a hard-decision denoiser with respect to the points in the constellation. Other widely-used single-step linear detectors include the matched filter and the \gls{mmse} detectors~\cite{massivemimobook} with $\bold{A}_0 = \bold{H}^H$ and $\bold{A}_0 = (\Hb^H \Hb+\sigma^2\bold{I}_{N_t})^{-1}\Hb^H$, respectively. Linear detectors are attractive for practical implementation because of their low complexity, but they perform substantially worse than the optimal detector.

We can also perform the optimization in \cref{eq:linear} in multiple iterations using gradient descent. The gradient of the objective function in the first equation of \cref{eq:linear} with respect to $\xb$ is $-2\Hb^H(\yb-\Hb\x)$. Hence, if we set $\textbf{A}_t$ to $2\alpha \Hb^H$ and $\textbf{b}_t=0$, the linear step of \cref{eq:general_fmwk} is equivalent to minimizing $||\y - \Hb \x||_2$ using gradient descent with step size $\alpha$. This is followed by a mapping onto the constellation set in the denoising step. If we had a compact convex constellation set, this projected gradient descent procedure is guaranteed to  converge to the global optimum. Discrete constellation sets, however, are not compact convex. Nonetheless, solving the linear least squares problem in \cref{eq:linear} iteratively may be desirable to avoid the cost of calculation the pseudo-inverse of the channel matrix.

\subsubsection{Approximate Message Passing (AMP)}\label{sec:amp} 
MIMO detection can, in principle, be solved through \gls{bp} if we consider a bipartite graph representation of the model in~\cref{eq:channel}~\cite{mezghani2010belief}. \gls{bp} on this graph requires $\mathcal{O}(N_rN_t)$ update messages in each iteration, which would be prohibitive for large system dimensions. In the large system limit,~\citet{jeon2015optimality} introduce \gls{amp} as a lower complexity inference algorithm for solving \cref{eq:problem} for \gls{iid} Gaussian channels. \gls{amp} reduces the number of messages in each iteration to $\mathcal{O}(N_r+N_t)$. The algorithm performs the following sequence of updates: 
\begin{align}\label{eq:amp}
\text{AMP:}&&
\begin{split}
    \textbf{z}_t &= \xh_{t} + \Hb^H (\y - \Hb\xh_t) +\textbf{b}_t\\
    \textbf{b}_t &= \alpha_t\left(\Hb^H(\yb-\Hb\xh_{t-1}) + \textbf{b}_{t-1}\right)\\
    \xh_{t+1} &=\eta_t\left( \textbf{z}_t; \sigma_t \right).
\end{split}
\end{align}
Consider AMP in our iterative framework: we use $\bold{A}_t=\Hb^H$ as the linear operator; the vector $\textbf{b}_t$ is known as the {\em Onsager} term; the scalar sequences $\sigma_t$ and $\alpha_t$ can be computed given the \gls{snr} and system parameters (constellation and number of transmitters and receivers)~\cite{bayati2011dynamics}. The denoising function $\eta_t(\cdot)$ applies the optimal denoiser for Gaussian noise in~\cref{eq:gaussian-denoiser} to each element of the vector $\bold{z}_t$. \citet{jeon2015optimality} prove that AMP is asymptotically optimal for large \gls{iid} Gaussian channel matrices. 

Orthogonal AMP (OAMP)~\cite{ma2017orthogonal} was proposed for unitarily-invariant channel matrices~\cite{tulino2004random} to relax the \gls{iid} Gaussian channel assumption in the original AMP algorithm:
\begin{align}\label{eq:oamp}
\text{OAMP:} &&
\begin{split}
    \textbf{z}_t &= \xh_{t} + \gamma_t \Hb^H{\left(v_t^2\Hb\Hb^H+\sigma^2 \textbf{I}\right)^{-1} } (\y - \Hb\xh_t)\\
    \xh_{t+1} &=\eta_t\left( \textbf{z}_t;\sigma_t^2 \right)\end{split}
\end{align}
where $\gamma_t={N_t}/{\text{trace}\left(v_t^2\Hb^H\left(v_t^2\Hb\Hb^H+\sigma^2 \textbf{I}\right)^{-1}\Hb\right)}$ is a normalizing factor and $v_t^2$ is proportional to the average noise power at the denoiser output at iteration $t$ and can be computed given the \gls{snr} and system dimensions. Notice that OAMP requires computing a matrix inverse in each iteration, making it more computationally expensive than AMP.

\subsubsection{Other techniques}
Several detection schemes relax the lattice constraint ($\textbf{x}\in \mathcal{X}^{N_t}$) in \cref{eq:problem}. For example, Semi-Definite Relaxation (SDR)~\cite{wiesel2005semidefinite} formulates the problem as a semi-definite program. Sphere decoding~\cite{viterbo1999universal} conducts a search over solutions $\xh$ such that $||\bold{y}-\bold{H\xh}||_2\le r$. Increasing $r$ covers a larger set of possible solutions, but this comes at the cost of increased complexity, approaching that of brute-force search. There is a large body of variations on improvements to this idea which can be found in \cite{yang2015fifty, larsson2009mimo}. While these approaches can perform well, their computational complexity is prohibitive for Massive MIMO systems with currently available hardware. 

Another class of detector applies several stages or iterations of linear detection followed by interference subtraction from the observation $\bold{y}$. The V-BLAST scheme \cite{wolniansky1998v} does this by  detecting the strongest symbols, which are then successively removed from  $\bold{y}$. The drawbacks of this approach are error propagation of early symbol decisions and high complexity due to the $N_t$ required stages, as well as the necessary reordering of transmitters after each step. Parallel interference cancellation (PIC) has been proposed to circumvent these problems. PIC jointly detects all transmitted symbols and then attempts to create an interference-free channel for each transmitter through the cancellation of all other transmitted symbols \cite{varanasi1990multistage, chin2002parallel}. A large system approximation of this approach was recently developed in \cite{shental2017massive} based on \cite{tanaka2005approximate}. However, it is currently limited to \gls{bpsk} modulation and leads to unsatisfactory performance for realistic system dimensions. 

In summary, most existing techniques are too complex to be implemented at the scale required by next-generation Massive MIMO systems. On the other hand, light-weight techniques like AMP cannot handle correlated channel matrices. These limitations have motivated a number of learning-based proposals for MIMO detection, which we discuss next.

\subsection{Learning-based MIMO detection schemes} \label{sec:mimo-learning}

\subsubsection{\detnet} \label{sec:DetNet}
Inspired by iterative projected gradient descent optimization, \citet{samuel2018learning} propose \detnet, a deep neural network architecture for MIMO detection. This architecture performs very well in case of \gls{iid} complex Gaussian channel matrices and achieves the performance of state-of-the-art algorithms for lower-order modulation schemes, such as \gls{bpsk} and QAM4. However, it is far more complex. The neural network is described by the following equations:
\begin{align}\label{eq:DetNet}
\text{DetNet:} &&
\begin{split}
\textbf{q}_t &= \xh_{t-1} - \theta^{(1)}_t \Hb^H\y + \theta^{(2)}_t\Hb^H\Hb\xh_{t-1}\\
\textbf{u}_t &= \left[\boldsymbol{\Theta}^{(3)}_t \textbf{q}_t + \boldsymbol{\Theta}^{(4)}_t \textbf{v}_{t-1} + \boldsymbol{\theta}^{(5)}_t\right]_+\\
\textbf{v}_t &= \boldsymbol{\Theta}^{(6)}_t \textbf{u}_t + \boldsymbol{\theta}^{(7)}_t\\
\xh_t &= \boldsymbol{\Theta}^{(8)}_t\textbf{u}_t + \boldsymbol{\theta}^{(9)}_t
\end{split}
\end{align} 
where $[x]_+=\max(x,0)$, which is also known as ReLU activation function~\cite{nair2010rectified}, is applied element-wise.

Although \detnet's performance is promising, it has two main limitations. First, its heuristic nature makes it difficult to reason about how the neural network works, and how to extend its architecture, for example, to support spatially correlated channel matrices or higher-order modulation schemes. Second, \detnet's architecture does not incorporate known properties of iterative methods and is thus unnecessarily complex. For example, many iterative soft-thresholding schemes (including AMP described above) apply a denoiser tailored for Gaussian noise \cref{eq:gaussian-denoiser} independently to each transmitted signal. \detnet's neural network can also be thought to be performing non-linear denoising steps intermixed with linear transformations. However, \detnet's denoisers are fully-connected 2-layer neural networks that operate on the entire vector of transmitted signals, i.e., they are $N_t$-dimensional functions instead of simple scalar functions.    

\subsubsection{OAMPNet} \citet{he2018model} design a learning-based iterative scheme based on the OAMP algorithm. OAMPNet adds two tuning parameters per iteration to the  OAMP algorithm. OAMPNet shows very good performance in the case of \gls{iid} Gaussian channels, but it does not generalize to realistic channels with spatial correlations, as our experiments in \cref{sec:experiments} show. The OAMPNet design can be expressed as: 
\begin{align}\label{eq:oampnet}
\text{OAMPNet:} &&
\begin{split}
    \textbf{z}_t &= \xh_{t} + \theta_t^{(1)}\Hb^H\left(v_t^2\Hb\Hb^H+\sigma^2 \textbf{I}\right)^{-1} (\y - \Hb\xh_t)\\
    \xh_{t+1} &=\eta_t\left( \textbf{z}_t;\sigma_t^2 \right).
\end{split}
\end{align}
OAMPNet uses the same denoisers used by AMP, which are optimal for Gaussian noise. 

By basing its design on OAMP, OAMPNet makes a strict assumption about the system: unitarily-invariant channel matrices. This reduces OAMPNet to training a few parameters per iteration. However, as our results show, OAMPNet's assumptions make it brittle, and its performance degrades on realistic channel matrices that do not conform to the assumptions. Further, like OAMP, OAMPNet must compute a matrix pseudo-inverse in each iteration and, therefore, its complexity is still quite high compared to schemes like AMP.
    \section{\sys Design}\label{sec:design}
The \sys design follows the iterative framework described in \Cref{sec:iter-framework}. The main idea behind \sys is to introduce the {\em right degree of flexibility} into the linear and denoising components of the iterative framework while preserving its overall structure. We observe that prior architectures do not strike the right balance between model flexibility and complexity.

In practice, not much is known about the distribution of channel matrices. We therefore propose a data-driven approach in which we learn a set of model parameters for each realization of $\Hb$. In this approach, the receiver continually adapts its parameters as it measures new channel matrices $\Hb$. We demonstrate that this can be realized in practice with a suitable neural network architecture by exploiting the fact that realistic channels exhibit locality in both the frequency and time domains. We introduce the neural network architecture in this section and discuss a practical training algorithm in \Cref{sec:cost}.

We present separate neural network models for (1) \gls{iid} Gaussian channels and (2) arbitrary channels. In the \gls{iid} Gaussian case, the model is extremely simple:
\begin{align}\label{eq:mmnet}
\text{\sys-iid:} &&
\begin{split}
        \textbf{z}_t &= \xh_{t} + \theta_t^{(1)}\Hb^H (\y - \Hb\xh_t)\\
        \xh_{t+1} &=\eta_t\left( \textbf{z}_t;{\sigma}_t^2 \right).
\end{split} 
\end{align}
Here, the denoiser is the optimal denoiser for Gaussian noise given in \eqref{eq:gaussian-denoiser}. \sysiid assumes the same distribution of noise at the input of the denoiser for all transmitted symbols and estimates its variance $\sigma_t^2$ according to
\begin{align}\label{eq:noise_est_iid}
\begin{split}
    {\sigma}_t^2 = \frac{{\theta}_t^{(2)}}{N_t}\left(\vphantom{\frac{\|\textbf{I}-\textbf{A}_t\Hb\|_F^2}{\|\Hb\|_F^2}} \right. &\left. \frac{\|\textbf{I}-\textbf{A}_t\Hb\|_F^2}{\|\Hb\|_F^2} \left[\|\y - \Hb\xh_{t}\|_2^2 - N_r \sigma^2\right]_+ \right. \\
    & + \left. \frac{\|\textbf{A}_t\|_F^2}{\|\Hb\|_F^2}\sigma^2 \right) .
\end{split}
\end{align}

The intuition behind \eqref{eq:noise_est_iid} is that the noise at the input of the denoiser at step $t$ is comprised of two parts: (i) the residual error caused by deviation of $\xh_t$ from the true value of $\x$,  and (ii) the contribution of the channel noise $\n$. The first component is amplified by the linear transformation $(\textbf{I} - \textbf{A}_t\Hb)$, and the second component is amplified by $\textbf{A}_t$. See~\cite{ma2017orthogonal, jeon2015optimality} for further details on this method for estimating noise variance. 

This model has only two parameters per layer: $\theta_t^{(1)}$ and $\theta_t^{(2)}$. We discuss this model merely to illustrate that, for the \gls{iid} Gaussian channel matrix case (which most prior work has used for evaluation), a simple model that adds a small amount of flexibility to existing algorithms like AMP can perform very well. In fact, our results will show that, in this case, we do not even need to train the parameters of the model online for each channel realization; training offline over randomly sampled \gls{iid} Gaussian channel suffices.

The \sys neural network for arbitrary channel matrices is as follows:
\begin{align}\label{eq:mmnet}
\text{\sys:} &&
\begin{split}
        \textbf{z}_t &= \xh_{t} + \boldsymbol{\Theta}_t^{(1)} (\y - \Hb\xh_t)\\
        \xh_{t+1} &=\eta_t\left( \textbf{z}_t;\boldsymbol{\sigma}_t^2 \right)
\end{split} 
\end{align}
where $\boldsymbol{\Theta}_t^{(1)}$ is an $N_t\times N_r$ complex-valued trainable matrix. In order to enable the model to handle cases in which different transmitted symbols have differing levels of noise, we add an extra degree of freedom to our estimations of noise per transmitter, resulting in:
\begin{align} \label{eq:noise_est}
\begin{split}
    \boldsymbol{\sigma}_t^2 = \frac{{\boldsymbol{\theta}}_t^{(2)}}{N_t}\left(\vphantom{\frac{\|\textbf{I}-\textbf{A}_t\Hb\|_F^2}{\|\Hb\|_F^2}}\right.  &\left.\frac{\|\textbf{I}-\textbf{A}_t\Hb\|_F^2}{\|\Hb\|_F^2} \left[\|\y - \Hb\xh_{t}\|_2^2 - N_r \sigma^2\right]_+ \right. \\
    &+ \left. \frac{\|\textbf{A}_t\|_F^2}{\|\Hb\|_F^2}\sigma^2 \right)
\end{split}
\end{align}
where the parameter vector $\boldsymbol{\theta}_t^{(2)}$  of size $N_t \times 1$ scales the noise variance by different amounts for each symbol. This approach distinguishes \sys from both the highly-constrained OAMPNet and overly complex DetNet solution. In particular, \sys uses a flexible linear transformation (which does not need to be linear in $\Hb$) to construct the intermediate signal $z_t$, but it applies the standard optimal denoiser for Gaussian noise in \eqref{eq:gaussian-denoiser}. Further, unlike OAMPNet, \sys does not require any expensive matrix inverse operation. 

\sys concatenates $T$ layers of the above form. We use the the average L2-loss over all $T$ layers in order to train the model, which is given by
\begin{equation}\label{eq:loss}
    L = \frac{1}{T}\sum_{t=1}^T \|\xh_t-\x\|_2^2.
\end{equation}
    \section{Experiments}\label{sec:experiments}
In this section, we evaluate and compare the performance of \sys with state-of-the-art schemes for both \gls{iid} Gaussian and realistic channel matrices. These are our main findings:
\begin{enumerate}
    \item
    On \gls{iid} Gaussian channels, most schemes perform very well. MMSE, SDR, V-BLAST and \detnet are 1-2dB far from the best schemes overall. AMP performance degrades for higher-order modulations in high SNRs. \sysiid and OAMPNet are very close to \opt in all experiment on these channels. \sysiid, however, has two orders of magnitude lower complexity than the learning-based schemes OAMPNet and \detnet.
    \item 
    On realistic, spatially-correlated channel matrices, the performance of all existing learning-based approaches degrades significantly. \sys ubiquitously shows the least gap with \opt. While \detnet and AMP fail to extend to these channels with reasonable performance (on QAM4, for example), MMSE has an 8--10dB gap with \opt on $64\times16$ channels. OAMPNet reduces this gap to 5--7dB. However, \sys closes the gap to less than 1.5dB. 
\end{enumerate}
Our implementations and channels dataset are available at \href{https://github.com/mehrdadkhani/MMNet}{https://github.com/mehrdadkhani/MMNet}.
\subsection{Methodology}
We first briefly discuss the details of detection schemes used for comparison. Since some of these schemes (including \sys) require training, we then discuss the process of generating data and training/testing on this data.  
\subsubsection{Compared Schemes}
In our experiments, we compare the following schemes on QAM modulation:
\begin{itemize}
\item {\bf MMSE:} Linear decoder that applies the SNR-regularized channel's pseudo inverse and rounds the output to the closest point on the constellation (see \cref{sec:linear}). 
\item {\bf SDR:} Semidefinite programming using a rank-1 relaxation interior point method~\cite{ma2008some}.  
\item {\bf V-BLAST:} Multi-stage interference cancellation BLAST algorithm using Zero-Forcing as the detection stage introduced in \cite{chin2002parallel}. 
\item {\bf AMP:} AMP algorithm for MIMO detection from~\citet{jeon2015optimality}. AMP runs 50 iterations of the updates as discussed in \eqref{eq:amp}. We verified that adding more iterations does not improve the results.
\item {\bf DetNet:} The deep learning approach introduced in \cite{samuel2018learning}. The DetNet paper describes instantiations of the architecture for BPSK, QAM4 and QAM16; these neural networks have, on the order of 1--10M, trainable parameters depending on the size of the system and constellation set. 
\item {\bf OAMPNet:} The OAMP-based architecture~\cite{he2018model} implemented in 10 layers with 2 trainable parameters per layer and an inverse matrix computation at each layer. 
\item {\bf \sysiid:} The simple architecture described in \Cref{sec:design}. This scheme has only $2$ scalar parameters per layer and does not require any matrix inversions. We implement this neural network with 10 layers.
\item {\bf \sys:} Our design described in \eqref{eq:mmnet}. It has 10 blocks, and the total number of trainable parameters is $2N_t(N_r+1)$ real values, independent of constellation size. 
\item {\bf \opt:} The optimal solver for \eqref{eq:problem} using a highly-optimized Mixed Integer Programming package Gurobi~\cite{gurobi}.
\end{itemize}

\subsubsection{Dataset}\label{sec:dataset}
Training and test data are generated through the model described in \eqref{eq:channel}. In this model, there are three sources of randomness: the signal $\xb$, the channel noise $\n$ and the channel matrix $\Hb$. Each transmitted signal $\xb$ is generated randomly and uniformly over the corresponding constellation set. All transmitters are assumed to use the same modulation. The channel noise $\n$ is sampled from a zero-mean \gls{iid} normal distribution with a variance that is set according to the operating \gls{snr}, defined as $\mathrm{SNR(dB)} = 10\log \left({\mathbb{E}[\|\Hb\x\|_2^2]}/{\mathbb{E}[\|\n\|_2^2]}\right)$. For every training batch, the \gls{snr}(dB) is chosen uniformly at random in the desired operating \gls{snr} interval. This interval depends on the modulation scheme. For each modulation in each experiment, the SNR regime is chosen such that the best scheme other than \opt can achieve a \gls{ser} of $10^{-3}$--$10^{-2}$. 

The channel matrices $\Hb$ are either sampled from an \gls{iid} Gaussian distribution (i.e., each column of $\Hb$ is a  complex-normal $\mathcal{CN}(0,(1/N_r)\bold{I}_{N_r})$),  or they are generated via the realistic channel simulation described below.

Here, we study two system size ratios ($N_t/N_r$) of 0.25 and 0.5, with the total number of receivers fixed at $N_r=64$. These are typical values for 4G/5G base stations in urban cellular deployments.  
For the case of realistic channels, we generate a dataset of channel realizations from the 3GPP 3D MIMO channel model \cite{3gpp-3d-mimo}, as implemented in the QuaDRiGa channel simulator \cite{jaeckel2014quadriga}.\footnote{The simulation results were generated using QuaDRiGa Version 2.0.0-664.} We consider a base station (BS) equipped with a rectangular planar array consisting of 32 dual-polarized antennas installed at a height of $25$\,m. The BS is assumed to cover a $120^\circ$-cell sector of radius $500\,$m within which $N_t\in\{16,32\}$ single-antenna users are dropped randomly. A guard distance of $10\,$m from the BS is kept. Each user is then assumed to move along a linear trajectory with a speed of $1\,$m/s. Channels are sampled every $\lambda/4\,$m at a center frequency of $2.53\,$GHz to obtain sequences of length 100. Each channel realization is then converted to the frequency domain assuming a bandwidth of $20\,$MHz and using $1024$ sub-carriers from which only every fourth is kept, resulting in $F=256$ effective sub-carriers. We gather a total of 40 user drops, resulting in $40\times 256$ length 100 sequences of channel matrices (i.e., 1M channel matrices in total).
Since the pathloss can vary dramatically between different users, we assume perfect power control, which normalizes the average received power across antennas and sub-carriers to one. Denote $\mathbf{H}[f,k]$ as the $k$th column of $\Hb$ on sub-carrier $f$. Our normalization ensures that 
$$ \frac{1}{N_r N_t F} \sum_{k=1}^{N_t}\sum_{f=1}^F \lVert\mathbf{H}[f,k]\rVert^2  = 1.$$

\subsubsection{Training} \sys, \detnet, and OAMPNet require training and were implemented in TensorFlow~\cite{tensorflow2015-whitepaper}. We have converted \cref{eq:problem} to its equivalent real-valued representation  for TensorFlow implementations (cf.\@ \cite[Sec.~II]{ma2009equivalence}). \detnet and OAMPNet are both trained as described in the corresponding publications (i.e., batch size of 5K samples). We trained each of the latter two algorithms for 50K iterations. 

To train \sys, we use the Adam optimizer~\cite{kingma2014adam} with a learning rate of $10^{-3}$. Each training batch has a size of 500 samples. We train \sys for 1K iterations on each realization of $\Hb$ in the naive implementation. In \Cref{sec:cost_reduction}, we exploit frequency and time domain correlations to reduce the training requirement to $4$ iterations per channel matrix. 

In spatially correlated channels, we do an additional 5K iterations of training with a batch size of 5K samples for each realization of $\Hb$ on the pre-trained OAMPNet model in order to have a fair comparison with \sys online training. However, as this extra training does not improve the performance of OAMPNet meaningfully, we do not count this re-training overhead in the complexity of OAMPNet algorithm in \cref{sec:cost}. 

For \gls{iid} Gaussian channels, \sysiid is not trained per channel realization $\Hb$. Instead, we use 10K iterations with a batch size of 500 samples to train a single \sysiid neural network, which we then test on new channel samples.

\subsection{Results}
We compare different schemes along two axes: performance and complexity. In this section, we focus on performance, leaving a comparison of complexity to \Cref{sec:cost}. We use the \gls{snr} required to achieve an \gls{ser} of $10^{-3}$ as the primary performance metric. In practice, most error correcting schemes operate around an \gls{ser} of $10^{-3}$--$10^{-2}$, so this is the relevant regime for MIMO detection. 

\subsubsection{\gls{iid} Gaussian channels}\label{sec:iid}
\begin{figure*}
    \centering
    \begin{subfigure}{\linewidth}
        \begin{tikzpicture}[smooth]
\begin{semilogyaxis}[
	xlabel={SNR(dB)},    
	ylabel={SER}, grid={both},width=1\linewidth,
        height=6cm, font=\small, xmin=3.8, xmax = 23.2, ymax=0.3, ymin=2e-5, legend style={at={(0.5,1.15)},fill=white, fill opacity=0.6, draw opacity=1,text opacity=1, draw=none, anchor=north,legend columns=8}, xtick distance=1, every axis plot/.append style={thick}, ylabel near ticks
]
\addplot coordinates {(4.0, 0.041034690539042185) (5.0, 0.027395516633987427) (6.0, 0.016846875349680657) (7.0, 0.00943593780199703) (8.0, 0.004699790477752597) (9.0, 0.001997288068135483)};

\addplot coordinates{(4,0.020921875000000)
    (5, 0.010578125000000)
    (6, 0.004515625000000)
   (7, 0.001528125000000)
   (8, 0.000396875000000)
   (9, 0.000078125000000)};
   
\addplot+ coordinates{
(4.000000, 0.037734)
(5.000000, 0.020589)
(6.000000, 0.009417)
(7.000000, 0.003608)
(8.000000, 0.000966)
(9.000000, 0.000222)
    };

\addplot coordinates {(4.0, 0.020446874999999975) (5.0, 0.009353124999999962) 
(6.0, 0.003782500000000022)
(7.0, 0.0013199999999999878)
(8.0, 0.0004028124999999605)
(9.0, 0.00011812499999996895)};

\addplot coordinates {(4.0, 0.02327687541643786) (5.0, 0.011755208174387688) (6.0, 0.005098539590835616) (7.0, 0.0018528103828431286) (8.0, 0.00057208140691134) (9.0, 0.00013510982195530374)};

\addplot coordinates{(4.0, 0.02005718946456947) (5.0, 0.00925229509671499) (6.0, 0.003576562404632777) (7.0, 0.0011929206053415964) (8.0, 0.00034114678700769563) (9.0, 8.239746093763323e-05)};

\addplot coordinates {(4.0, 0.020148120323816743) (5.0, 0.009259998798370361) (6.0, 0.003611461321512799) (7.0, 0.0011725087960562375) (8.0, 0.0003210365772244872) (9.0, 7.083415985120745e-05)};

\addplot coordinates {(4.0, 0.02026249999999996) (5.0, 0.00895593750000001)
(6.0, 0.0034287499999999804)
(7.0, 0.0010868750000000427)
(8.0, 0.0002784375000000061)
(9.0, 6.249999999996536e-05)};

\pgfplotsset{cycle list shift=-8}
\addplot coordinates {
(11.0, 0.0784944812456766) (12.0, 0.052802815039952766) (13.0, 0.03274468580881762) (14.0, 0.01834062337875353) (15.0, 0.009223232666651482) (16.0, 0.00392406384150179) 
};

\addplot coordinates{
(11.0, 0.05193)
(12.0, 0.03067)
(13.0, 0.01623)
(14.0, 0.00687)
(15.0, 0.0025)
(16.0, 0.00061)
};

\addplot+ coordinates{
(11.000000, 0.073563)
(12.000000, 0.042392)
(13.000000, 0.020175)
(14.000000, 0.007542)
(15.000000, 0.002177)
(16.000000, 0.000478)
    };

\addplot coordinates {(11.0, 0.04223750000000004)
(12.0, 0.019334375000000015) (13.0, 0.007221875000000044) (14.0, 0.0026906249999999465) (15.0, 0.0008531250000000101)
(16.0, 0.00032812500000001243)
  };

\addplot coordinates {(11.0, 0.050717397530873676) (12.0, 0.027187709013620776) (13.0, 0.012386357784271218) (14.0, 0.004834997653961004) (15.0, 0.0015834410985311465) (16.0, 0.000454370180765884)};

\addplot+ coordinates{(11.0, 0.04052635312080377) (12.0, 0.01788385192553188) (13.0, 0.0064126014709477985) (14.0, 0.0018905206521356854) (15.0, 0.0004727081457773785) (16.0, 9.479482968599573e-05)};

\addplot coordinates {(11.0, 0.04225364724795) (12.0, 0.019262198607126968) (13.0, 0.007054791847864728) (14.0, 0.0021099984645841507) (15.0, 0.0005389551321665076) (16.0, 0.00011281172434463027)};

\addplot+ coordinates{
(11, 0.0419688)
(12, 0.0173125)
(13, 0.005725)
(14, 0.00189375)
(15, 0.00046875)
(16, 0.000096875)
};
\pgfplotsset{cycle list shift=-16}
\addplot coordinates {(18.0, 0.07136114637057) (19.0, 0.04518437782923379) (20.0, 0.026029372215270796) (21.0, 0.013260418176651023) (22.0, 0.005949264764785633) (23.0, 0.002245203653971295)
};

\addplot coordinates{
    (18, 0.0548)
    (19, 0.0316)
    (20, 0.0163)
    (21, 0.0067)
    (22, 0.0025)
    (23, 7.7187e-04)};

\addplot coordinates{
(18.000000, 0.064602)
(19.000000, 0.032331)
(20.000000, 0.013383)
(21.000000, 0.004123)
(22.000000, 0.000892)
(23.000000, 0.000122)
     };
\addplot coordinates {
(18.0, 0.03494937499999995)
(19.0, 0.013642187500000014)
(20.0, 0.005146562500000007)
(21.0, 0.0018681249999999983)
(22.0, 0.0009550000000000392)
(23.0, 0.000634687499999953)};

\addplot coordinates {(18.0, 0.06303156018257139) (19.0, 0.03910926977793361) (20.0, 0.02216364542643212) (21.0, 0.011388854185739983) (22.0, 0.005283127228418949) (23.0, 0.002234059572220004)};

\addplot coordinates{(18.0, 0.03347166736920637) (19.0, 0.011901353597640996) (20.0, 0.00353593985239653) (21.0, 0.0009141651789348337) (22.0, 0.00020916541417392853) (23.0, 4.1453043619688046e-05)};

\addplot coordinates {(18.0, 0.03768010934193933) (19.0, 0.015162913004557521) (20.0, 0.0047157287597658915) (21.0, 0.001223444938659557) (22.0, 0.00028656721115116746) (23.0, 5.8956940968801774e-05)
};
\addplot coordinates {(18.0, 0.02828125) (19.0, 0.010375) (20.0, 0.00303125)
(21.0, 0.000705625000000043) (22.0, 0.00014843750000004263)
 (23.0, 2.312500000001272e-05)};

\node (source) at (axis cs:4,0.15){};
       \node (destination) at (axis cs:10,0.15){};
       \draw[<->, thick](source)--(destination);
\node (source) at (axis cs:10,0.15){};
       \node (destination) at (axis cs:17,0.15){};
       \draw[<->, thick](source)--(destination);
\node (source) at (axis cs:17,0.15){};
       \node (destination) at (axis cs:23,0.15){};
       \draw[<->, thick](source)--(destination);
\node at (axis cs:7,0.15) [anchor=center, font=\small, fill=white] {QAM4};
\node at (axis cs:13.5,0.15) [anchor=center, font=\small, fill=white] {QAM16};
\node at (axis cs:20,0.15) [anchor=center, font=\small, fill=white] {QAM64};

\legend{MMSE, SDR, V-BLAST, AMP, \detnet, OAMPNet, \sysiid, \opt} 

\end{semilogyaxis}
\end{tikzpicture}
        \caption{32 Transmit, 64 Receive Antennas}
         \label{fig:iid_Nt32}
    \end{subfigure}
    \begin{subfigure}{\linewidth}
        \begin{tikzpicture}[smooth] 
\begin{semilogyaxis}[
	xlabel={SNR(dB)},    
	ylabel={SER}, grid={both},width=1\linewidth,
        height=6cm, font=\small, xmin=1.8, xmax = 21.2, ymax=0.08, ymin=6e-5, legend style={at={(0.5,1.2)},fill=white, fill opacity=0.6, draw opacity=1,text opacity=1, draw=none, anchor=north,legend columns=8}, xtick distance=1, every axis plot/.append style={thick}, ylabel near ticks]
\addplot+  table[x=x,y=mean] {figures/SERvsSNR_mmse_M4_iid.dat};

\addplot+ coordinates{
    (2, 0.0075)
    (3, 0.0033)
    (4, 0.0010)
    (5, 3.3750e-04)
    (6, 8.5000e-05)
    };

\addplot+  table[x=x,y=mean] {figures/SERvsSNR_BLAST_M4_iid.dat};

\addplot+  table[x=x,y=mean] {figures/SERvsSNR_AMP_M4_iid.dat};

\addplot+  table[x=x,y=mean] {figures/SERvsSNR_lin_DetNet_M4_iid.dat};

\addplot+  table[x=x,y=mean] {figures/SERvsSNR_lin_oamp_M4_iid.dat};

\addplot+  table[x=x,y=mean] {figures/SERvsSNR_aHt_M4_iid.dat};

\addplot+  table[x=x,y=mean] {figures/SERvsSNR_ML_M4_iid.dat};
\pgfplotsset{cycle list shift=-8}

\addplot+  table[x=x,y=mean] {figures/SERvsSNR_mmse_M16_iid.dat};

\addplot+ coordinates{
(9, 0.0163)
(10, 0.0072)
(11, 0.0028)
(12, 7.1250e-04)
(13, 1.8125e-04) 
(14, 5.6250e-05)
};

\addplot+  table[x=x,y=mean] {figures/SERvsSNR_BLAST_M16_iid.dat};

\addplot+  table[x=x,y=mean] {figures/SERvsSNR_AMP_M16_iid.dat};

\addplot+  table[x=x,y=mean] {figures/SERvsSNR_lin_DetNet_M16_iid.dat};

\addplot+  table[x=x,y=mean] {figures/SERvsSNR_lin_oamp_M16_iid.dat};

\addplot+  table[x=x,y=mean] {figures/SERvsSNR_aHt_M16_iid.dat};

\addplot+  table[x=x,y=mean] {figures/SERvsSNR_ML_M16_iid.dat};

\pgfplotsset{cycle list shift=-16}
\addplot+  table[x=x,y=mean] {figures/SERvsSNR_mmse_M64_iid.dat};

\addplot+ coordinates{
(16, 0.0130)
(17, 0.0056)
(18, 0.0019)
(19, 5.5625e-04)
(20, 1.1875e-04)
(21, 6.2500e-06)
};

\addplot+  table[x=x,y=mean] {figures/SERvsSNR_BLAST_M64_iid.dat};

\addplot+  table[x=x,y=mean] {figures/SERvsSNR_AMP_M64_iid.dat};

\addplot+  table[x=x,y=mean] {figures/SERvsSNR_lin_DetNet_M64_iid.dat};

\addplot+  table[x=x,y=mean] {figures/SERvsSNR_lin_oamp_M64_iid.dat};

\addplot+  table[x=x,y=mean] {figures/SERvsSNR_aHt_M64_iid.dat};

\addplot+  table[x=x,y=mean] {figures/SERvsSNR_ML_M64_iid.dat};

\legend{MMSE, SDR, V-BLAST, AMP, \detnet, OAMPNet, \sysiid, \opt} 

\node (source) at (axis cs:2,0.05){};
       \node (destination) at (axis cs:8,0.05){};
       \draw[<->, thick](source)--(destination);
\node (source) at (axis cs:8,0.05){};
       \node (destination) at (axis cs:15,0.05){};
       \draw[<->, thick](source)--(destination);
\node (source) at (axis cs:15,0.05){};
       \node (destination) at (axis cs:21,0.05){};
       \draw[<->, thick](source)--(destination);
\node at (axis cs:5,0.05) [anchor=center, font=\small, fill=white] {QAM4};
\node at (axis cs:11.5,0.05) [anchor=center, font=\small, fill=white] {QAM16};
\node at (axis cs:18,0.05) [anchor=center, font=\small, fill=white] {QAM64};
\end{semilogyaxis}
\end{tikzpicture}
        \caption{16 Transmit, 64 Receive Antennas} 
        \label{fig:iid_Nt16}
    \end{subfigure}
    \caption{\gls{ser} vs. \gls{snr} of different schemes for three modulations (QAM4, QAM16 and QAM64) and two system sizes (32 and 16 transmitters, 64 receivers) with \gls{iid} Gaussian channels.}
    \label{fig:iid}
\end{figure*}
\Cref{fig:iid} shows the \gls{ser} vs. \gls{snr} of the state-of-the-art MIMO schemes on \gls{iid} channels for two system sizes: 32 and 16 transmitters (\cref{fig:iid_Nt32} and \cref{fig:iid_Nt16}, respectively). 

We make the following observations:
\begin{enumerate}
    \item The \gls{snr} required to achieve a certain \gls{ser} increases by 2--3dB as we double the number of transmitters (notice the different range of x-axes). The reason is that there is higher interference with more transmitters.
  
    \item
    There is a 2--3dB performance gap between \opt and MMSE across all modulations for $N_t=32$. However, this gap decreases to 1dB for $N_t=16$, because received signals have lower interference in this case. 
    
    \item
     Multiple schemes perform similarly to \opt, especially at lower-order modulations like QAM4. As we move to QAM64, the performance of several schemes degrades compared to \opt. 

    \item
    SDR performs better than MMSE, but its gap with \opt increases with modulation order. 
    
    \item
    V-BLAST achieves almost the optimal performance across all modulations when we have 16 transmit antennas. However, its performance is sensitive to system size and degrades when we increase the number of transmitters to 32.
    
    \item
    AMP is near-optimal in many cases (recall that, theoretically, AMP is asymptotically optimal for \gls{iid} Gaussian channels). However, it suffers from robustness issues at higher SNR levels, especially with higher-order modulations like QAM64. 
    
    \item
    \detnet has a good performance on QAM4, but its gap with \opt increases as we move to QAM16 and QAM64. With QAM64, \detnet performs even worse than MMSE for $N_t=16$. 
    
    \item 
    OAMPNet and the simple \sysiid approach are both very close to \opt across different modulations over a wide range of SNRs, even though these models have only two parameters per layer. Unlike OAMPNet, \sysiid does not require any matrix inversions. As we discuss in \Cref{sec:cost}, \sysiid has $\mathcal{O}(N_r)\times$ lower computational complexity than OAMPNet because it does not need matrix inversions and must learn only 20 parameters in total, compared to the more than 1M trainable parameters of \detnet.

\end{enumerate}

\begin{figure*}[htb!]
    \centering
    \begin{subfigure}{\linewidth}
    \begin{tikzpicture}[smooth] 
\begin{semilogyaxis}[
	xlabel={SNR(dB)},    
	ylabel={SER}, grid={both},width=1\linewidth,
        height=6cm, font=\small, xmin=7.8, xmax = 30.2, ymax=0.8, ymin=6e-5, legend style={at={(0.5,1.2)},fill=white, fill opacity=0.6, draw opacity=1,text opacity=1, draw=none, anchor=north,legend columns=4}, xtick distance=1, every axis plot/.append style={thick}, ylabel near ticks]
\addplot+ [thick] table[x=x,y=mean] {figures/Nt_32/SERvsSNR_mmse_M4_correlated.dat};

\addplot+ [thick] table[x=x,y=mean] {figures/Nt_32/SERvsSNR_lin_oamp_M4_correlated.dat};

\addplot+ [thick] table[x=x,y=mean] {figures/Nt_32/SERvsSNR_fixed_W_M4_correlated.dat};

\addplot+ [thick] table[x=x,y=mean] {figures/SERvsSNR_ML_M4_correlated_Nt32.dat};

\pgfplotsset{cycle list shift=-4} 
\addplot+ [thick] table[x=x,y=mean] {figures/Nt_32/SERvsSNR_mmse_M16_correlated.dat};

\addplot+ [thick] table[x=x,y=mean] {figures/Nt_32/SERvsSNR_lin_oamp_M16_correlated.dat};

\addplot+ [thick] table[x=x,y=mean] {figures/Nt_32/SERvsSNR_fixed_W_M16_correlated.dat};

\addplot+ [thick] table[x=x,y=mean] {figures/SERvsSNR_ML_M16_correlated_Nt32.dat};

\pgfplotsset{cycle list shift=-8} 
\addplot+ [thick] table[x=x,y=mean] {figures/Nt_32/SERvsSNR_mmse_M64_correlated.dat};

\addplot+ [thick] table[x=x,y=mean] {figures/Nt_32/SERvsSNR_lin_oamp_M64_correlated.dat};

\addplot+ [thick] table[x=x,y=mean] {figures/Nt_32/SERvsSNR_fixed_W_M64_correlated.dat};
\addplot+ [thick] table[x=x,y=mean] {figures/SERvsSNR_ML_M64_correlated_Nt32.dat};

\legend{MMSE, OAMPNet, \sys, \opt}
\node (source) at (axis cs:8,0.5){};
       \node (destination) at (axis cs:17,0.5){};
       \draw[<->, thick](source)--(destination);
\node (source) at (axis cs:17,0.5){};
       \node (destination) at (axis cs:24,0.5){};
       \draw[<->, thick](source)--(destination);
\node (source) at (axis cs:24,0.5){};
       \node (destination) at (axis cs:30,0.5){};
       \draw[<->, thick](source)--(destination);
\node at (axis cs:12.5,0.5) [anchor=center, font=\small, fill=white] {QAM4};
\node at (axis cs:20.5,0.5) [anchor=center, font=\small, fill=white] {QAM16};
\node at (axis cs:27,0.5) [anchor=center, font=\small, fill=white] {QAM64};
\end{semilogyaxis}
\end{tikzpicture}
    \caption{32 Transmit, 64 Receive Antennas}
    \label{fig:correlated_Nt32}
    \end{subfigure}
    \begin{subfigure}{\linewidth}
    \begin{tikzpicture}[smooth] 
\begin{semilogyaxis}[
	xlabel={SNR(dB)},    
	ylabel={SER}, grid={both},width=1\linewidth,
        height=6cm, font=\small, xmin=3.8, xmax = 25.2, ymax=0.8, ymin=6e-5, legend style={at={(0.5,1.2)},fill=white, fill opacity=0.6, draw opacity=1,text opacity=1, draw=none, anchor=north,legend columns=4}, xtick distance=1, every axis plot/.append style={thick}, ylabel near ticks]
\addplot+ [thick] table[x=x,y=mean] {figures/SERvsSNR_mmse_M4_correlated.dat};

\addplot+ [thick] table[x=x,y=mean] {figures/SERvsSNR_lin_oamp_M4_correlated.dat};

\addplot+ [thick] table[x=x,y=mean] {figures/SERvsSNR_fixed_W_M4_correlated.dat};

\addplot+ [thick] table[x=x,y=mean] {figures/SERvsSNR_ML_M4_correlated_Nt16.dat};
\pgfplotsset{cycle list shift=-4} 
\addplot+ [thick] table[x=x,y=mean] {figures/SERvsSNR_mmse_M16_correlated.dat};

\addplot+ [thick] table[x=x,y=mean] {figures/SERvsSNR_lin_oamp_M16_correlated.dat};

\addplot+ [thick] table[x=x,y=mean] {figures/SERvsSNR_fixed_W_M16_correlated.dat};
\addplot+ [thick] table[x=x,y=mean] {figures/SERvsSNR_ML_M16_correlated_Nt16.dat};
\pgfplotsset{cycle list shift=-8} 
\addplot+ [thick] table[x=x,y=mean] {figures/SERvsSNR_mmse_M64_correlated.dat};

\addplot+ [thick] table[x=x,y=mean] {figures/SERvsSNR_lin_oamp_M64_correlated.dat};

\addplot+ [thick] table[x=x,y=mean] {figures/SERvsSNR_fixed_W_M64_correlated.dat};

\addplot+ [thick] table[x=x,y=mean] {figures/SERvsSNR_ML_M64_correlated_Nt16.dat};
\legend{MMSE, OAMPNet, \sys, \opt}
\node (source) at (axis cs:4,0.5){};
       \node (destination) at (axis cs:12,0.5){};
       \draw[<->, thick](source)--(destination);
\node (source) at (axis cs:12,0.5){};
       \node (destination) at (axis cs:19,0.5){};
       \draw[<->, thick](source)--(destination);
\node (source) at (axis cs:19,0.5){};
       \node (destination) at (axis cs:25,0.5){};
       \draw[<->, thick](source)--(destination);
\node at (axis cs:8,0.5) [anchor=center, font=\small, fill=white] {QAM4};
\node at (axis cs:15.5,0.5) [anchor=center, font=\small, fill=white] {QAM16};
\node at (axis cs:22,0.5) [anchor=center, font=\small, fill=white] {QAM64};
\end{semilogyaxis}
\end{tikzpicture}
    \caption{16 Transmit, 64 Receive Antennas}
    \label{fig:correlated_Nt16}
    \end{subfigure}
    \caption{\gls{ser} vs. \gls{snr} of different schemes for three modulations (QAM4, QAM16 and QAM64) and two system sizes (32 and 16 transmitters, 64 receivers) with \real channels.}
    \label{fig:correlated}
\end{figure*}
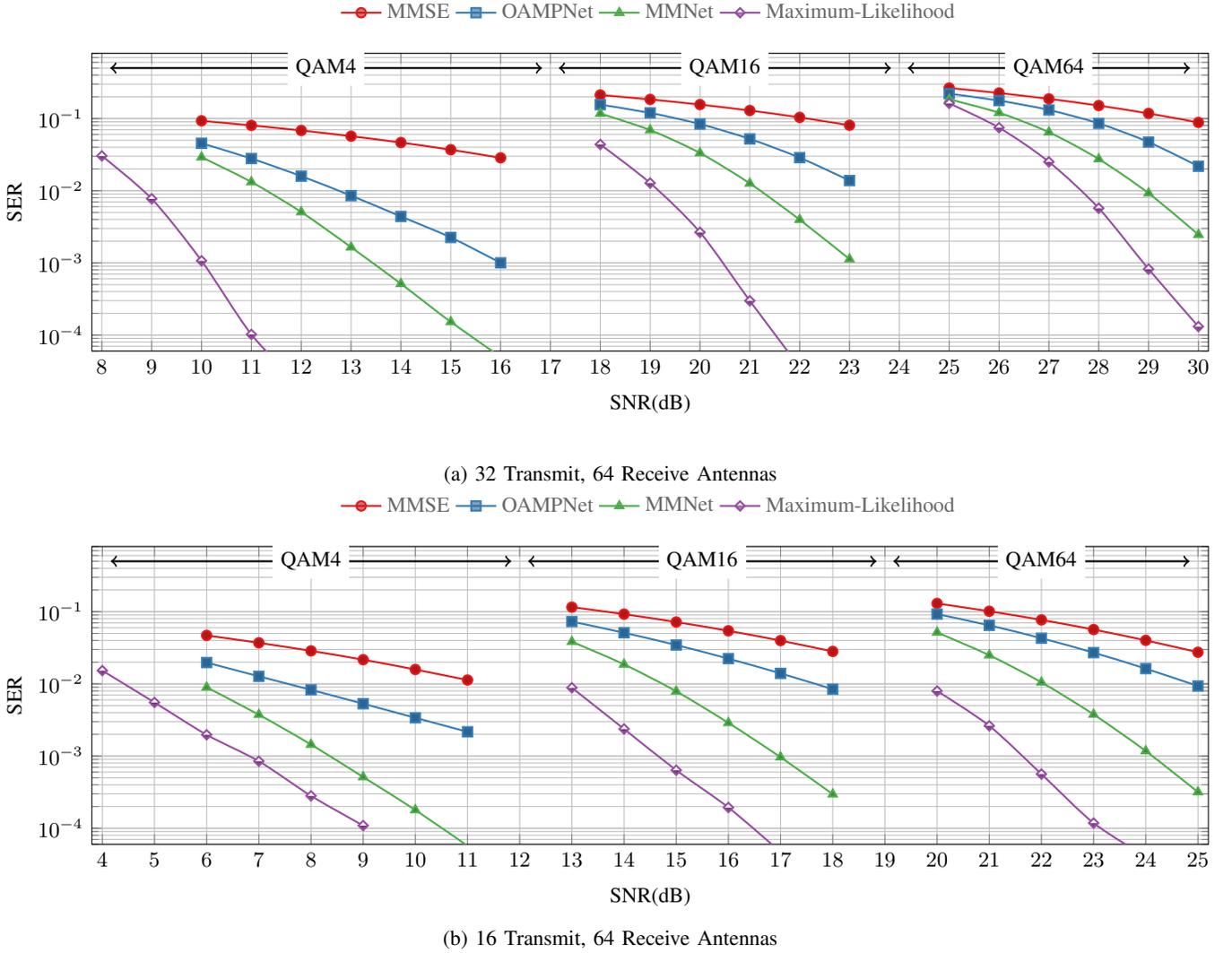

In summary, these results show that for \gls{iid} Gaussian channel matrices, adding just a small amount of flexibility via tuning parameters to existing iterative schemes like \gls{amp} can result in equivalent or improved performance over much more complex deep learning models like \detnet. These models may even outperform classical algorithms like \gls{sdr}.

\subsubsection{Realistic channels}

\Cref{fig:correlated} shows the results for the realistic channels derived using the 3GPP 3D MIMO channel model. We consider only MMSE (as a baseline), OAMPNet, \sys and \opt. As shown in the \gls{iid} Gaussian case, schemes like SDR, V-BLAST and \detnet do not perform as well as the OAMPNet baseline.\footnote{We tried to train \detnet for realistic channels and ran into significant difficulty with stability and convergence in training.} Also, AMP is not designed for correlated channels and is known to perform poorly (see discussion in~\cref{sec:error_analysis}).   

We make the following observations:
\begin{enumerate}
    \item 
    There is a much larger gap with \opt for all detection schemes on these channels compared to the \gls{iid} case. 
    \item 
    We require 4--7dB increase in the SNR ranges relative to \cref{fig:iid}. Also, doubling the number of transmitters from \cref{fig:correlated_Nt16} to \cref{fig:correlated_Nt32} costs about 5dB in SNR for each scheme this time (compare with 2--3dB in \gls{iid} case.)
    \item 
    MMSE as a baseline shows a relatively flat SER vs. SNR in this case. For example, it requires 5dB SNR improvement on QAM16 to go from an SER of $2\%$ to $1\%$.
    \item
    OAMPNet performance improvement slope is faster than MMSE. It shows 2--3dB average improvement in SNR requirement relative to MMSE to achieve the same SER.
    \item 
    \sys outperforms MMSE and OAMNet schemes for both system sizes and in all modulations. 
\end{enumerate}

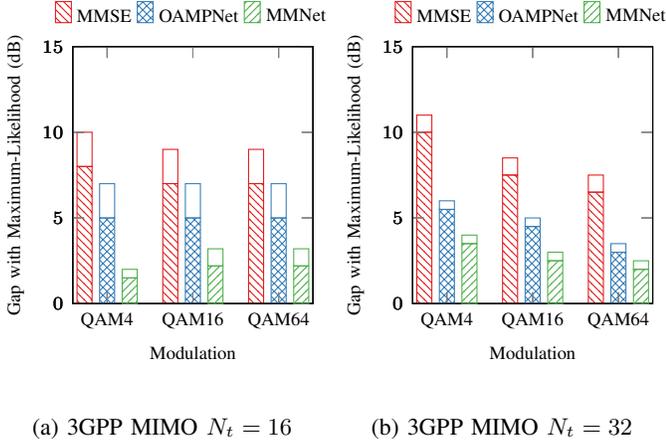
\begin{figure}
\centering
\begin{subfigure}[b]{0.49\linewidth}
\begin{tikzpicture}
\begin{axis}[
    ybar stacked, 
    draw = none, 
    bar width = 2mm,
    bar shift = -3mm,
    cycle multi list=Set1-9,
    every axis plot/.append style={fill},
    ytick scale label code/.code={},
    ymin=0.,
    ymax=15,
    legend style={at={(0.5,1.2)}, draw=none,fill=none,
      anchor=north,legend columns=-1},
    symbolic x coords={QAM4,QAM16,QAM64},
    xlabel = {Modulation},
    ylabel = {Gap with \opt (dB)},
    xtick=data,
    font=\scriptsize,
    height=5cm,
    width=1.1\linewidth,
    enlarge x limits=0.2,
    ylabel near ticks,
    xlabel near ticks
    ]
      \addlegendimage{forget plot, color=Set1-A, pattern color=.,pattern=north west lines};
  \addlegendentry{MMSE};
  \addlegendimage{forget plot, color=Set1-B, pattern color=.,pattern=crosshatch};
  \addlegendentry{OAMPNet};
  \addlegendimage{forget plot, color=Set1-C, pattern color=.,pattern=north east lines};
  \addlegendentry{\sys};
\addplot+[forget plot, pattern color=.,pattern=north west lines] coordinates {(QAM4, 8) (QAM16, 7) (QAM64, 7)};
\addplot+[fill=none] coordinates {(QAM4,2) (QAM16,2) (QAM64, 2)};
\end{axis}
\begin{axis}[
    ybar stacked, 
    bar width = 2mm,
    bar shift= 0mm,
    cycle multi list=Set1-9,
    ytick scale label code/.code={},
    every axis plot/.append style={fill},
    ymin=0.,
    ymax=15,
    legend style={at={(5cm,3cm)}, draw=none,fill=none,
      anchor=north,legend columns=-1},
    symbolic x coords={{},{{}}, {{{}}}},
    xtick=data,
    font=\scriptsize,
    height=5cm,
    width=1.1\linewidth,
    enlarge x limits=0.2,
    ylabel near ticks,
    xlabel near ticks
    ]
\pgfplotsset{cycle list shift=-1}
\addplot+[forget plot, pattern color=.,pattern=crosshatch] coordinates {({},5) ({{}},5) ({{{}}}, 5)};
\addplot+[fill=none] coordinates {({},2) ({{}},2) ({{{}}}, 2)};
\end{axis}
\begin{axis}[
    ybar stacked, 
    bar width = 2mm,
    bar shift= 3mm,
    cycle multi list=Set1-9,
    every axis plot/.append style={fill},
    ytick scale label code/.code={},
    ymin=0.,
    ymax=15,
    legend style={at={(47mm,2.5cm)}, draw=none,fill=none,
      anchor=north,legend columns=-1},
    symbolic x coords={{},{{}}, {{{}}} },
    xtick=data,
    font=\scriptsize,
    height=5cm,
    width=1.1\linewidth,
    enlarge x limits=0.2,
    ylabel near ticks,
    xlabel near ticks
    ]
\pgfplotsset{cycle list shift=-2}
\addplot+[forget plot, pattern color=.,pattern=north east lines] coordinates {({}, 1.5) ({{}}, 2.2) ({{{}}}, 2.2)};
\addplot+[fill=none] coordinates {({},0.5) ({{}},1) ({{{}}}, 1)};

\end{axis}
\end{tikzpicture}
\caption{\real $N_t=16$}
\end{subfigure}
\hfill
\begin{subfigure}[b]{0.49\linewidth}
\begin{tikzpicture}
\begin{axis}[
    ybar stacked, 
    draw = none, 
    bar width = 2mm,
    bar shift = -3mm,
    cycle multi list=Set1-9,
    every axis plot/.append style={fill},
    ytick scale label code/.code={},
    ymin=0.,
    ymax=15,
    legend style={at={(0.5,1.2)}, draw=none,fill=none,
      anchor=north,legend columns=-1},
    symbolic x coords={QAM4,QAM16,QAM64},
    xlabel = {Modulation},
    ylabel = {Gap with \opt (dB)},
    xtick=data,
    font=\scriptsize,
    height=5cm,
    width=1.1\linewidth,
    enlarge x limits=0.2,
    ylabel near ticks,
    xlabel near ticks
    ]
          \addlegendimage{forget plot, color=Set1-A, pattern color=.,pattern=north west lines};
  \addlegendentry{MMSE};
  \addlegendimage{forget plot, color=Set1-B, pattern color=.,pattern=crosshatch};
  \addlegendentry{OAMPNet};
  \addlegendimage{forget plot, color=Set1-C, pattern color=.,pattern=north east lines};
  \addlegendentry{\sys};
\addplot+[forget plot, pattern color=.,pattern=north west lines] coordinates {(QAM4, 10) (QAM16, 7.5) (QAM64, 6.5)};
\addplot+[fill=none] coordinates {(QAM4,1) (QAM16,1) (QAM64, 1)};
\end{axis}
\begin{axis}[
    ybar stacked, 
    bar width = 2mm,
    bar shift= 0mm,
    cycle multi list=Set1-9,
    ytick scale label code/.code={},
    every axis plot/.append style={fill},
    ymin=0.,
    ymax=15,
    legend style={at={(5cm,3cm)}, draw=none,fill=none,
      anchor=north,legend columns=-1},
    symbolic x coords={{},{{}}, {{{}}}},
    xtick=data,
    font=\scriptsize,
    height=5cm,
    width=1.1\linewidth,
    enlarge x limits=0.2,
    ylabel near ticks,
    xlabel near ticks
    ]
\pgfplotsset{cycle list shift=-1}
\addplot+[forget plot, pattern color=.,pattern=crosshatch] coordinates {({},5.5) ({{}},4.5) ({{{}}}, 3)};
\addplot+[fill=none] coordinates {({},0.5) ({{}},.5) ({{{}}}, 0.5)};
\end{axis}
\begin{axis}[
    ybar stacked, 
    bar width = 2mm,
    bar shift= 3mm,
    cycle multi list=Set1-9,
    every axis plot/.append style={fill},
    ytick scale label code/.code={},
    ymin=0.,
    ymax=15,
    legend style={at={(47mm,2.5cm)}, draw=none,fill=none,
      anchor=north,legend columns=-1},
    symbolic x coords={{},{{}}, {{{}}} },
    xtick=data,
    font=\scriptsize,
    height=5cm,
    width=1.1\linewidth,
    enlarge x limits=0.2,
    ylabel near ticks,
    xlabel near ticks
    ]
\pgfplotsset{cycle list shift=-2}
\addplot+[forget plot, pattern color=.,pattern=north east lines] coordinates {({}, 3.5) ({{}}, 2.5) ({{{}}}, 2)};
\addplot+[fill=none] coordinates {({},0.5) ({{}},0.5) ({{{}}}, 0.5)};
\end{axis}
\end{tikzpicture}
\caption{\real $N_t=32$}
\end{subfigure}
\caption{SNR requirement gap with \opt at SER of $10^{-3}$. The total bar height shows the 90th-percentile gap (over different channels) while the hatched section depicts the average.}
\label{fig:perf_gap}
\end{figure}

In \cref{fig:perf_gap}, we plot the performance gap with \opt for these three detection schemes. For this purpose, we measure the difference in the minimum SNR level that is required to have SER of $10^{-3}$. In this figure, we have also attempted to show the variability of this requirement across different channel situations by adding the 90th-percentile gap over different channel conditions. We observe that \sys can achieve up to 5dB and 8dB improvement, respectively, over OAMPNet and MMSE on more realistic channels.
    \section{Why \sys works}\label{sec:error_analysis}
In this section, we examine why \sys performs better than other schemes. By analyzing the dynamics of the error ($\xh_t - \x$), we find that \sys's denoisers are significantly more effective than those in OAMPNet. We show that this occurs because \sys's linear stages control the distribution of noise at the input of the denoisers. Specifically, \sys ensures that the noise input to the denoisers is nearly Gaussian, whereas the noise distribution for OAMPNet is far from Gaussian. Since the denoisers in both architectures are tailored for Gaussian noise, they perform much better in \sys. 
\subsection{Error dynamics}

Define the error at the outputs of the linear and denoiser stages at iteration $t$ as $\textbf{e}_t^{lin} = \textbf{z}_t - \x$ and $\textbf{e}_t^{den}=\xh_{t+1} - \x$, respectively. For algorithms such as \sys and OAMPNet with $\textbf{b}_t = 0$,  we can rewrite the update equations of \eqref{eq:general_fmwk} in terms of these two errors in the form:
\begin{subequations}\label{eq:err_dynamics}
\begin{align}
         \textbf{e}_t^{lin} &= (\textbf{I}-\textbf{A}_t\Hb)  \textbf{e}_{t-1}^{den} + \textbf{A}_t \n \label{eq:err_dynamics_lin}\\
         \textbf{e}_{t}^{den} &=\eta_t( \x + \textbf{e}_t^{lin} ) - \x. \label{eq:err_dynamics_nonlin}
\end{align}
\end{subequations}
\Cref{eq:err_dynamics_lin} includes two terms, corresponding to the contribution of the error of the previous stages' output and the channel noise to $\elin_t$ respectively. To gain intuition, consider the effect of several choices of $\textbf{A}_t$. 

If we set $\textbf{A}_t$ to $\Hb^+$ (the pseudo-inverse of the channel matrix), we are only left with the term $\Hb^+\n$ in $\elin_t$, thus eliminating the error caused by the previous stage entirely. However, this comes at a price: we are left with Gaussian noise with covariance matrix $\sigma^2\Hb^+{\Hb^+}^H$. This presents two potential problems: (i) if $\Hb$ is ill-conditioned, it might amplify the channel noise (e.g., inversely proportional to the smallest singular value of $\Hb$ in some directions); (ii) if $\elin_t$ is correlated noise, applying an element-wise denoising function to it may not be effective. 
    
We could also remove the channel noise term entirely by setting $\textbf{A}_t$ to zero. This would effectively remove the linear stage. However, if optimal denoisers are used, removing the linear stage and applying the denoiser function twice should have no effect on reducing the error.

For \gls{iid} Gaussian channels with variance $1/N_r$, if we set $\bold{A}_t=\Hb^H$, the factor $\textbf{I} - \bold{A}_t\Hb$ asymptotically goes to zero as we increase $N_r$~\cite{bayati2011dynamics}. The auto-covariance of $\bold{A}_t\n$, $\sigma^2\Hb^H\Hb$, is approximately equal to $\sigma^2\bold{I}_{N_t}$. This means that the channel noise term is neither amplified nor correlated following this linear transformation with \gls{iid} channels, while the error from the previous stage, $\eden_{t-1}$, is attenuated significantly via $\textbf{I} - \bold{A}_t\Hb$. These calculations explain why AMP has great performance on \gls{iid} Gaussian channels. However, in case of correlated channels, neither $\textbf{I} - \bold{A}_t\Hb$ is close to zero, nor is $\bold{A}_t\n$ uncorrelated. This is the primary reason that AMP cannot perform well on more realistic channels. 

\subsection{Analysis}
\begin{figure}
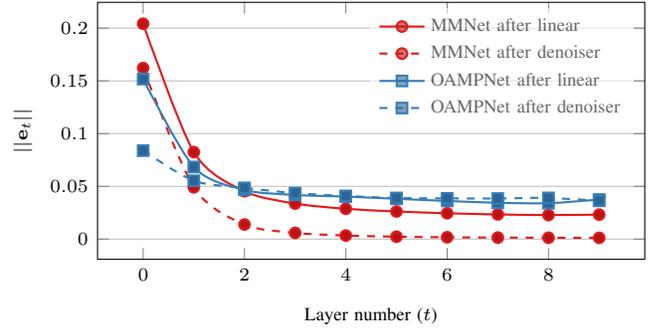

    \centering
    \begin{tikzpicture}[smooth]
        \begin{axis}[ylabel=$||\bold{e}_t||$, xlabel=Layer number ($t$), font=\scriptsize, ymajorgrids=true, legend cell align={left}, legend style={fill=white, fill opacity=0.6, draw opacity=1,text opacity=1,draw=none},yticklabel style={/pgf/number format/fixed,
                  /pgf/number format/precision=2}, width=\linewidth, height=5cm, ylabel near ticks, ytick distance=0.05]
            \input{figures/rawfiles/inputerrorvslayer_learnW.tex}
            \input{figures/rawfiles/errorvslayer_learnW.tex}
            \input{figures/rawfiles/inputerrorvslayer_oampW.tex}
            \input{figures/rawfiles/errorvslayer_oampW.tex}
            \legend{\sys after linear, \sys after denoiser, OAMPNet after linear, OAMPNet after denoiser};
        \end{axis}
    \end{tikzpicture}
    \caption{Noise power after the linear and denoiser stages at different layers of OAMPNet and \sys. The OAMPNet denoisers become ineffective after the third layer on \real channels.}
    \label{fig:input_vs_output_err}
\end{figure}
 As noted earlier, the key element in \sys and prior schemes such as OAMPNet is how to pick the linear transformation $\bold{A}_t$. Based on the above discussion of the error dynamics, we identify two desirable properties:
\begin{enumerate}
    \item
    $\bold{A}_t$ must reduce the magnitude of $\elin_t$. This requires striking a {\em balance} between the two terms in \eqref{eq:err_dynamics_lin}, because attenuating one term may amplify the other and vice-versa. 
    
    \item
    $\bold{A}_t$ must ``shape'' the distribution of $\elin_t$ to make it suitable for the subsequent denoiser. In particular, the denoisers in most iterative schemes (e.g., \sys and OMAPNet) are specifically designed for \gls{iid} Gaussian noise. Thus, ideally, the linear stage should avoid outputting correlated or non-Gaussian noise.
    
\end{enumerate}

\Cref{fig:input_vs_output_err} shows the average noise power at the output of the linear and denoiser stages across iterations, for both OAMPNet and \sys on $64\times16$ \real channels. The average noise power before and after the denoiser saturate at the same value in OAMPNet from the third layer ($t=2$) onwards, showing that the denoisers are no longer effective after a few iterations.  

We hypothesize that the reason OAMPNet's denoisers become ineffective is that the noise distribution for OAMPNet is not Gaussian, whereas \sys is able to provide near-Gaussian noise to its denoiser. We evaluate how close the noise distribution is to Gaussian for both schemes using the Anderson test~\cite{anderson1952asymptotic}. In order to measure this score, we generate 10,000 samples per channel realization $\Hb$. We then calculate the Anderson score for the noise distribution at each linear stage output per transmitter, and per channel matrix. If this score is below a threshold of 0.786, it indicates that the noise comes from a Gaussian distribution with a significance of $5\%$, i.e. the probability of false rejection of a Gaussian distribution is less than $5\%$.

\begin{figure}
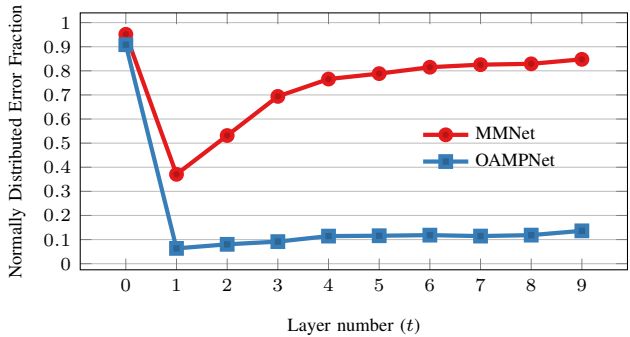

\centering
\begin{tikzpicture}
\begin{axis}[legend style={at={(0.75,0.6)}, draw=none, fill=none,
      anchor=north,legend columns=1},legend cell align={left}, xlabel={Layer number ($t$)}, ylabel={Normally Distributed Error Fraction},ymajorgrids=true, width=\linewidth, height=5cm, font=\scriptsize, ylabel near ticks, xtick distance=1, ytick distance=0.1]
\input{figures/rawfiles/anderson_score_cdf/hist_summary_learnW.tex}
\input{figures/rawfiles/anderson_score_cdf/hist_summary_oampW.tex}
\legend{\sys, OAMPNet};
\end{axis}
\end{tikzpicture}
\caption{Percentage of transmitters that have Gaussian error distribution after the linear block for each layer with significance level of $5\%$. \sys produces Normal-distributed errors at the output of linear blocks, while OAMPNet fails to achieve the Gaussian property.}
\label{fig:linear_anderson}
\end{figure}
In \cref{fig:linear_anderson}, we plot the average ratio of transmitters that have Normally-distributed noise at the output of the linear stage according to this test. Since in both schemes we start with $\xh_0=0$, the output of the linear stage at layer $t=0$ is $\textbf{A}_0\n$, which is Gaussian. Thus, the fraction of transmitters with Gaussian noise is 1 in layer $t=0$ for both schemes. However, both schemes deviate from Gaussian noise in layer $t=1$ while sharply reducing the total noise power as seen in \cref{fig:input_vs_output_err}. However, \sys deviates less from a Gaussian distribution. Unlike OAMPNet, in which the noise for 95\% of the transmitters is not Gaussian at layer $t=1$, for \sys nearly 40\% of the transmitter exhibit Gaussian noise. On the other hand, \sys reduces the noise power slightly less than OAMPNet in layer $t=1$. 

In subsequent layers, the noise distribution for \sys becomes increasingly Gaussian, with nearly 90\% of transmitter passing the Anderson test by layer $t=9$. By contrast, most transmitters in OAMPNet continue to exhibit non-Gaussian noise in subsequent layers, though the fraction of transmitter with Gaussian noise increases marginally.   

Next, we measure the effect of input error power on the output error distribution of linear stages in both schemes. In other words, we are interested to know how $\|\eden_{t-1}\|$ impacts the Gaussian distribution property of $\elin_t$. For this purpose, we choose the median of Anderson scores as a measure of the linear stage's ability to control the distribution of its output errors. In \cref{fig:andersonvserror}, we show the 2D histogram of this median score for different values of $\|\eden_{t-1}\|$. We also plot three thresholds of $1\%$, $5\%$ and $15\%$ significance for the normality test with dashed horizontal lines as a reference. To be Normally distributed with 1\%, 5\%, or 15\% confidence, the Anderson scores must fall below the respective line.

We notice that the median score in both schemes increases with the norm of the error from the previous denoiser stage. In other words, the linear stages that have a higher input noise power produce outputs that deviate more from Gaussian noise.  However, \sys is $100\times$ better in terms of the median score at controlling the input error for large value of $\|\eden_{t-1}\|$. This figure also suggests that the poor performance of OAMPNet in final layers is likely not only because of the aggressive approach it has taken at $t=1$. Later linear stages are also not very good at controlling the distribution of their output errors.

\subsection{Impact of channel condition number}
\begin{figure}
\pgfplotsset{
colormap={grayred}{color(0cm)=(black!10); color(1cm)=(Set1-A!75!black)}
}
\begin{subfigure}[t]{0.49\linewidth}
\begin{tikzpicture}[]
\begin{axis}[
    xtick = {-3,-2,-1,0},
    xticklabels={$10^{-3}$,$10^{-2}$,$10^{-1}$,$10^0$},
    ytick = {-1, 0, 1, 2},
    yticklabels = {$10^{-1}$, $10^0$, $10^1$, $10^2$},
    xlabel = {$\|\eden_{t-1}\|$},
    ylabel = {median(Anderson Score($\elin_t$))},
    ylabel near ticks,
    enlargelimits=false,
    colorbar, colormap name=grayred,
           scale only axis,width=\linewidth, unit vector ratio*=1 1 1,
           colorbar horizontal,
        colorbar style={
            font = \scriptsize,
            title=count in bin,
            at={(0,1.0)},               
            anchor=below south west,    
            width=\pgfkeysvalueof{/pgfplots/parent axis width},
        },
        colorbar/width=2.5mm,
clip mode=individual,
]
\addplot+ [
    scatter, scatter/use mapped color={draw=mapped color, fill=mapped color},
    scatter src=explicit,
    only marks,
    mark=square*,
] table [x=x, y=y ,col sep=space, meta=z] {./figures/rawfiles/func_hist_oampW.csv};
\addplot+[mark=none, black, thick, dashed] coordinates {(-3,-0.2395775166) (-0.15,-0.2395775166)};
\addplot+[mark=none, black, thick, dashed] coordinates {(-3,-0.104577454) (-0.15,-0.104577454)};
\addplot+[mark=none, black, thick, dashed] coordinates {(-3,0.03782475059) (-0.15,0.03782475059)};
\end{axis}
\end{tikzpicture}
\caption{OAMPNet}
\end{subfigure}
\begin{subfigure}[t]{0.49\linewidth}
\begin{tikzpicture}[]
\begin{axis}[
    xtick = {-3,-2,-1,0},
    xticklabels={$10^{-3}$,$10^{-2}$,$10^{-1}$,$10^0$},
    ytick = {-1, 0, 1, 2},
    yticklabels = {$10^{-1}$, $10^0$, $10^1$, $10^2$},
    xlabel = {$\|\eden_{t-1}\|$},
    ylabel = {median(Anderson Score($\elin_t$))},
    ylabel near ticks,
    enlargelimits=false,
    colorbar, colormap name=grayred,
           scale only axis,width=\linewidth, unit vector ratio*=1 1 1,
           colorbar horizontal,
           colorbar style={
            font = \scriptsize,
            title=count in bin,
            at={(0,1.0)},               
            anchor=below south west,    
            width=\pgfkeysvalueof{/pgfplots/parent axis width},
        },
        colorbar/width=2.5mm,
clip mode=individual,
]
\addplot+ [
    scatter, scatter/use mapped color={draw=mapped color, fill=mapped color},
    scatter src=explicit,
    only marks,
    mark=square*,
] table [x=x, y=y ,col sep=space, meta=z] {./figures/rawfiles/func_hist_learnW.csv};
\addplot+[mark=none, black, thick, dashed] coordinates {(-3,-0.2395775166) (-0.15,-0.2395775166)};
\addplot+[mark=none, black, thick, dashed] coordinates {(-3,-0.104577454) (-0.15,-0.104577454)};
\addplot+[mark=none, black, thick, dashed] coordinates {(-3,0.03782475059) (-0.15,0.03782475059)};
\end{axis}
\end{tikzpicture}
\caption{\sys}
\end{subfigure}
\caption{Median of Anderson score for the noise at the output of linear stage vs. the power of noise at the input of the stage. \sys controls the linear block output error distribution to be Gaussian. Dashed horizontal lines show the thresholds for $1\%$, $5\%$ and $15$ significance level.}
\label{fig:andersonvserror}
\end{figure}
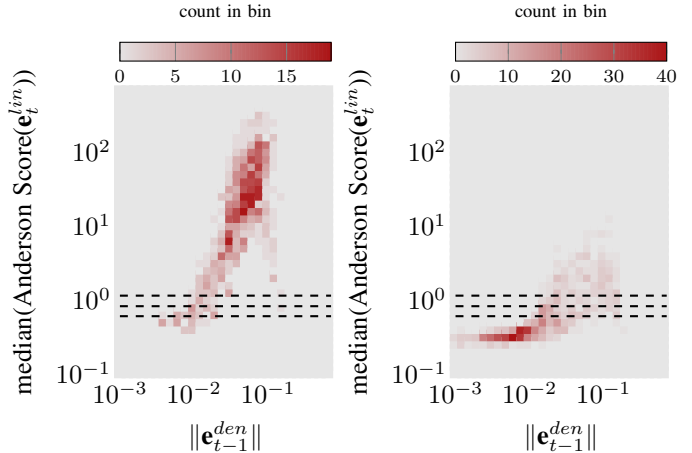
\begin{figure*}[t!]
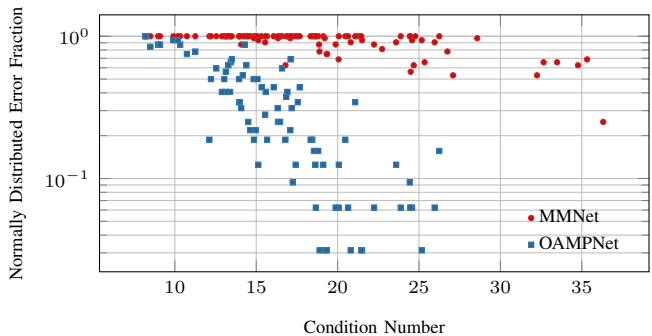
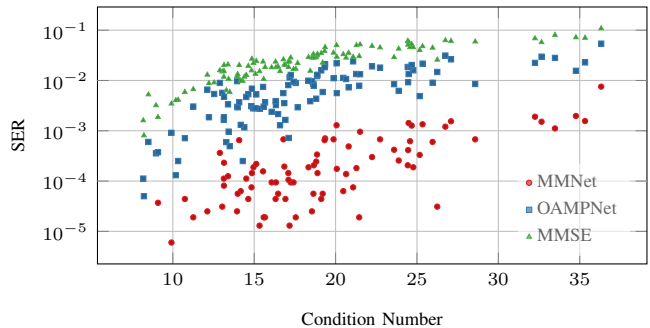

\begin{subfigure}[t]{0.49\linewidth}
\begin{tikzpicture}[every mark/.append style={mark size=1pt}]
    \begin{semilogyaxis}[xlabel=Condition Number, grid=both, ylabel={Normally Distributed Error Fraction}, ylabel near ticks, font=\scriptsize, width=\linewidth, height=5cm, legend pos = south east,legend style={draw=none, fill=none}, legend cell align={left}, xtick distance=5]
        \input{figures/rawfiles/Gaussianratiovscond/Gaussianratiovscond_learnW_4.tex}
        \input{figures/rawfiles/Gaussianratiovscond/Gaussianratiovscond_oampW_4.tex}
        \legend{\sys, OAMPNet}
    \end{semilogyaxis}
\end{tikzpicture}
\caption{Ratio of $\elin$ Gaussians vs. channel condition number }
\label{fig:cond_effect_ratio}
\end{subfigure}
\hfill
\begin{subfigure}[t]{0.49\linewidth}
\begin{tikzpicture}[every mark/.append style={mark size=1pt}]
    \begin{semilogyaxis}[width=1.1\linewidth, height=5cm, grid={both}, xlabel=Condition Number, ylabel=SER, ylabel near ticks, font=\scriptsize, width=\linewidth, legend pos = south east,legend style={fill=white, fill opacity=0.6, draw opacity=1,text opacity=1, draw=none}, legend cell align={left}, extra y ticks={1e-2,1e-4}, xtick distance=5]
    \input{figures/rawfiles/accvscond_learnW.tex}
    \input{figures/rawfiles/accvscond_oampW.tex}
    \input{figures/rawfiles/accvscond_mmse.tex}
    \legend{\sys, OAMPNet, MMSE};
    \end{semilogyaxis}
\end{tikzpicture}
\caption{SER on \real channels}
\label{fig:cond_effect_SER_real}
\end{subfigure}

\caption{Effect of condition number on performance of schemes. (a) \sys is more robust in producing the right noise distribution for denoisers with changes in condition number. (b) SER is directly affected by the condition number. }
\label{fig:cond_effect}
\end{figure*}
Finally, we evauate the impact of the channel {\em condition number} on \sys and OAMPNet. A channel's condition number is defined as the ratio of its largest singular value to the smallest. It is well-known that symbol detection is difficult for  channel matrices with higher condition number.

OAMPNet tries to address this issue by introducing filtering over the singular values. In fact, the linear update equation in \eqref{eq:oampnet} attempts to map each singular value $s$ in $\Hb$ to $\theta_t^{(1)}s/(v_t^2s^2+\sigma^2)$. This in turn attempts to dynamically adjust the shape of the sphere mapping in each iteration by tuning $\theta_t^{(1)}$ and $v_t^2$. We see that if all singular values are near each other, as is usually the case in \gls{iid} Gaussian channels, OAMPNet easily maps each sphere of signals in the transmit space to a non-skewed sphere in the receive signals space. However, our results show that manipulating the singular values is not the best option for poorer channel condition numbers.

In \cref{fig:cond_effect_ratio}, we show the scatterplot of performance of \sys linear stage in terms of preserving the Gaussian distribution property of their output error distribution across different channel condition numbers sample from our \real dataset after a few initial iterations ($t=4$). We see that the fraction of Normally-distributed errors decreases quickly for OAMPNet with the increase in condition number, while \sys maintains the ratio for a broader range of condition numbers in \real channels. The consequence of this failure in meeting the underlying assumptions of the model shows up in \cref{fig:cond_effect_SER_real}. In this figure, we show the scatterplot of SER at different condition numbers for \real channels. Although all schemes' performances are affected by the condition number, \sys can almost maintain an SER of around $10^{-3}$ almost across all conditions. 
    \section{Online Training Algorithm}\label{sec:cost}
Training deep-learning models is a computationally intensive task, often requiring hours or even days for large models. The computation overhead depends on two factors: (i) the total number of required training samples, and (ii) the size of the model. For example, in training a model like \detnet with an order of 1M parameters, we need 50K iterations with a batch size of 5K samples, i.e., 250M training samples. If we assume each parameter of the model shows up in at least one floating-point operation per training sample, we require a minimum of $2.5\times10^{14}$ floating-point operations for the entire process of training \detnet. This extreme computational complexity makes training such models online for each realization of $\bold{H}$  impossible. 

By contrast, \sys has only $\sim$40K parameters, and training it from scratch requires about 1000 iterations with batch size of 500. Further, this section will show that by taking advantage of locality of the channels observed at a receiver, we can effectively train the model for each channel realization in four iterations (with batch size 500) on average. Thus training \sys has six orders of magnitude lower computational overhead than \detnet, making online training for each realization of $\Hb$ practical. 

In this section, we first discuss the temporal and spectral locality of \real channels. Next, we show how we can exploit these localities to accelerate online training.

\subsection{Channel locality}
The channel matrices measured at a base station exhibit two forms of locality: 
\begin{itemize}
    \item {\em Temporal:} Channel matrices change gradually over time as user devices move within a cell or the multipath environment changes. The samples of $\Hb$ at nearby points in time are thus correlated. 
    
    \item {\em Spectral}: A \gls{bs} needs to recover signals from several frequency subcarriers (1024 in our \real model). The channels for subcarriers in nearby frequencies are also strongly correlated, as the frequency merely affects the phase for multipath signal components incident on receiver antennas. For a path of length $l$, the phase difference for two subcarriers $\Delta f$ apart in frequency is $\propto l\cdot\Delta f$. Therefore the received signal and the channels for adjacent subcarriers will be similar at each receiver antenna.  

\end{itemize}

Both forms of locality reduce the complexity of training for each channel realization, because (i) the cost of channel-specific computations can be amortized across multiple correlated channel realizations across time and frequency, (ii) the trained model for one channel realization can serve as strong initialization for training for adjacent channel realizations in the time-frequency plane.  
    
\begin{figure}
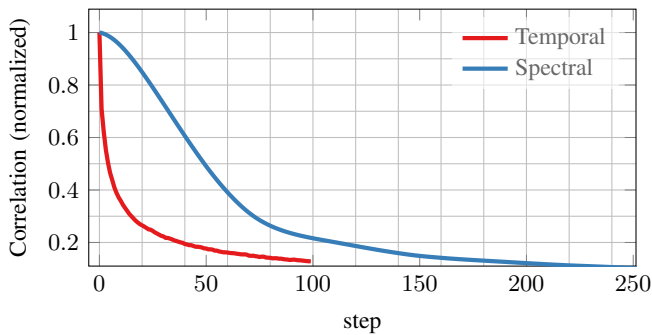

    \begin{tikzpicture}
\begin{axis}[
	xlabel={step},
	ylabel={Correlation (normalized)}, grid=none,xminorgrids=true, yminorgrids=true, width=\linewidth,
        height=5cm, font=\small, ylabel near ticks, legend pos = north east,legend style={fill=white, fill opacity=0.6, draw opacity=1,text opacity=1, draw=none}, legend cell align={left}, xtick={0,50,...,300},
                             minor xtick={0,20,...,280}, minor tick style={draw=none}, xmin=-5,ymin=0.11, minor ytick={0.1,0.2,...,1}, ytick={0.2,0.4,...,1}]
    \input{figures/time_autocorrelation.tex}
    \input{figures/channel_autocorrelation.tex}
    \legend{Temporal, Spectral}
\end{axis}
\end{tikzpicture}
        \caption{Correlation of channel samples across time and frequency dimensions. The correlation decays relatively quickly in the time dimension, but the channel matrices show strong locality across sub-carriers in frequency dimension.}
        \label{fig:cor}
\end{figure}

\Cref{fig:cor} shows both forms of locality by plotting the correlation among the \real channel samples across time and frequency (subcarriers)
In order to compute these correlations, we have averaged the inner-product of the channel matrices with their shifted realization by the step value in the corresponding dimension. In this experiment, a shift of one step in the time domain corresponds to two channels at the same subcarrier frequency that are 118ms apart in time. Similarly, a shift of 1 step in the frequency domain corresponds to two channels at adjacent subcarriers (78.1KHz apart) at the same time. We normalize the inner-product by the norm of matrices, such that the correlation of a channel matrix with itself is 1. We observe a stronger locality in the frequency domain than in the time domain on these channels. 

\subsection{Training algorithm and results}\label{sec:cost_reduction}

In this section, we show how channel locality can help reduce the total number of operations \sys needs to decode each received signal $\yb$ at the \gls{bs}. The computational complexity of \sys is mostly dominated by the cost of online training for each new realization of the channel $\Hb$.  This cost in turn depends on the channel {\em coherence time}. In the case of a quasi-static channel, as expected for instance in fixed-wireless access or backhaul solutions such as 5G home wireless (see Section 7.6.2 in~\cite{massivemimobook}), the channel between the transmitter and receiver does not change for extended periods of time. In such cases, \sys does not require frequent retraining and can reuse the same model until the communication channel changes significantly. However, \sys can also operate at reasonable computation cost when the channel is changing fairly frequently. For example, our \real channel samples were generated assuming all devices constantly move at a speed of 1 m/s, and after about 500ms, the channel correlation is less than 0.5. However, even in this scenario, we require only 4 training iterations on average per channel realization, as explained next.

To see how, note that a receiver at a \gls{bs} must simultaneously decode signals from different subcarriers. Since channels exhibit strong correlations across sub-carriers (\cref{fig:cor}), training the \sys detector on $\Hb$ for one subcarrier produces a detector that will achieve very similar performance on adjacent subcarriers. The performance of this detector will however decay for more distant subcarriers in the frequency domain. 

Based on this observation, we propose the online training scheme in Algorithm \ref{alg:training}. We start from a random initialization of the \sys neural network model $\mathcal{M}$. We define $n$ as an index for time intervals in which we can assume that channels do not change substantially. For each interval $n$, we measure a channel matrix $\Hb[f]$ for each subcarrier frequency $f$. The basic idea in the algorithm is to train the model for 1000 iterations (with a batch size of 500) for the first subcarrier ($f = 1$), then retrain the model using only 3 additional training iterations per subcarrier for all subsequent subcarriers. For each channel matrix $\Hb[f]$, we generate $(\x,\y)$ training data pairs using \cref{eq:channel}. Once the model has trained for subcarrier $f$, we save a copy of the model as $\mathcal{M}_n[f]$ for detecting all signals received in time interval $n$ on that subcarrier. We repeat the entire training algorithm in each time interval. 

\begin{figure}[t]
    \centering
    \input{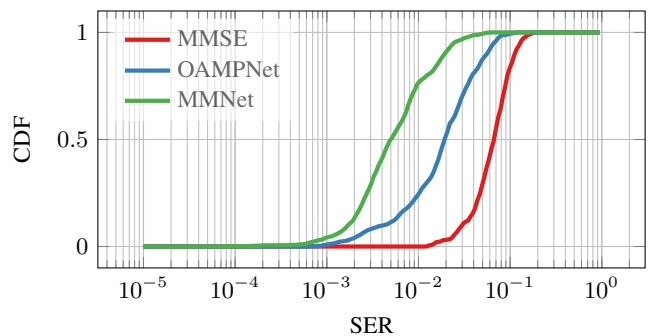}
    \caption{CDF of SER using Algorithm \ref{alg:training} for training \sys on QAM16 modulation. \sys requires only 4 overall iterations of batch size 500 per channel realization to train to a reasonable performance.}
    \label{fig:cdf}
\end{figure}

\begin{algorithm}[t]
  \caption{\sys online training} \label{alg:training}
  \small{
  \begin{algorithmic}
  \State $\mathcal{M}$ $\leftarrow$ model(\sys) \Comment load \sys model
  \State $\mathcal{M}$.initialize(\ ) \Comment{Initialize model parameters randomly}
  \State batchsize $\leftarrow$ 500 \Comment Training batch size
  \State $n \leftarrow 0$ \Comment{$n$ keeps time}
    \While  {True}
        \For {$f$ in $\{1,2,...,F\}$ }
            \State $\Hb[f]$ $\leftarrow$ Measured channel at time $n$ and frequency $f$
            \State $\textbf{D} $ $\leftarrow$ Generate random $(\x,\y)$ pairs for $\Hb[f]$ using \cref{eq:channel}
            \If {$f==1$}
                \State numTrainIterations $\leftarrow 1000$  
            \Else 
                \State numTrainIterations $\leftarrow 3$
            \EndIf
            \State $\mathcal{M}$.train(\textbf{D}, numTrainIterations, batchsize) \Comment{Update model parameters}
            \State $\mathcal{M}_{n}[f] \leftarrow \mathcal{M}$.copy(\ ) 
        \EndFor
        \State $n \leftarrow n+1$
    \EndWhile 
  \end{algorithmic}}
\end{algorithm}  
 In \Cref{fig:cdf}, we show the result of this online training method on \real channels for the QAM16 modulation. Roughly $95\%$ of samples are decoded with SER of less than $0.02$, while OAMPNet (trained explicitly for thousands of iterations on each realization of $\Hb$ at each time interval $n$ and subcarrier $f$) is only able to decode $50\%$ of the samples at the same error rate. With this approach, \sys performs 3.55K iterations of training with batch size 500 in order to learn a detector for all 1024 subcarriers in total at each time interval $n$. Therefore the cost of online training is less than 4 iterations on average per channel realization, yet \sys delivers better performance than other schemes, like OAMPNet and MMSE. 
 
\subsection{Computational complexity}

One iteration of training \sys on a batch of size of $b$ has a complexity of $\mathcal{O}(bN_r^2)$, as detection takes $\mathcal{O}(N_r^2)$ in \sys. To put this in perspective, a light-weight algorithm like AMP has a complexity of $\mathcal{O}(N_r^2)$ dominated by the multiplication of the channel matrix and residual vectors. The MMSE scheme has a higher complexity of $\mathcal{O}(N_r^3)$ because it needs to invert a matrix. OAMPNet similarly requires a matrix inversion, resulting in a complexity of $\mathcal{O}(N_r^3)$. 

\begin{figure}
\centering
\begin{tikzpicture}
    \begin{semilogyaxis}[
            ylabel = {\#Multiplication Ops.}, ymajorgrids=true,
            ybar,
            symbolic x coords={MMSE,AMP,\sys,\detnet,OAMPNet},
            xtick=data, width=\linewidth, height=6cm,
            font=\small,
            cycle multi list=Set1-9,
            every axis plot/.append style={fill},
            legend pos=north west,legend style={fill=white, fill opacity=0.6, draw opacity=1,text opacity=1, draw=none}, 
        ]
        \addplot+[pattern color=.,pattern=north east lines] coordinates{(MMSE, 5734.400000) (AMP, 40960.000000) (OAMPNet, 26255360.000000) (\sys, 1679360.000000) (\detnet, 3686400.000000)};
        \addplot+[pattern color=.,pattern=north west lines] coordinates{(MMSE, 16056.320000) (AMP, 81920.000000) (OAMPNet, 31539200.000000) (\sys, 3358720.000000) (\detnet, 14745600.000000)};
        \legend{$N_t=16$, $N_t=32$}
    \end{semilogyaxis}
\end{tikzpicture}
\caption{Number of multiplication operations per signal detection for different algorithms on QAM16 with $N_r=64$ receive antennas in \real model. Detection with \sys, including its online training process, requires fewer multiplication operations than detection with pre-trained \detnet and OAMPNet models.}
\label{fig:cost_comp}
\end{figure}
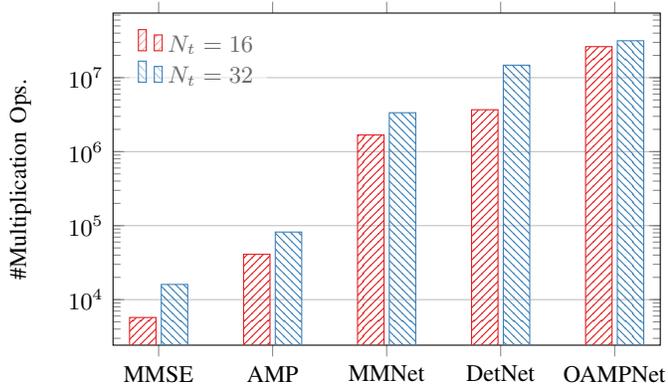

Moving beyond $\mathcal{O}(\cdot)$ analysis, \cref{fig:cost_comp} shows the average number of multiplication operations required per signal detection on \real channels for learning-based algorithms in addition to two classic baselines, MMSE and AMP. All algorithms may reuse their computation if feasible. In particular, in every interval of channel coherence in 3GPP MIMO model, each algorithm receives  $\sim$100 signals to detect~\cite[Definition 2.2 on page 220]{massivemimobook}. MMSE takes the best advantage of this as it calculates the required channel matrix inversion only once for all 100 received signals in the coherence interval. This way, MMSE depreciates its principal bottleneck calculation by a factor of 100$\times$ resulting in 5--7$\times$ less computation than AMP algorithm which cannot reuse the calculations but still has very moderate computation overhead by design. \sys with online training and detection operations together, places after AMP with 2--5$\times$ fewer multiplications than pre-trained \detnet. However, neither AMP nor \detnet extend to spatially correlated channels. \sys reuses the weights it trains with 4 iterations of batch size 500 for all 100 received signals in a coherence interval. However, unlike MMSE, OAMPNet has to calculate a new matrix inversion in each layer for every received signal as $v_t^2$ in \cref{eq:oampnet} depends on $\xh_t$.

Consequently, the cost of \sys with its online training algorithm is 10--15$\times$ less than OAMPNet depending on the system size. \sys has only 41$\times$ higher computational complexity than a very light iterative approach like \gls{amp}, which only works near \opt under specific circumstances of \gls{iid} Gaussian channels.

    \section{Conclusion and Future Work}
This paper proposed a new deep learning architecture for Massive MIMO detection with an online training algorithm. \sys outperforms state-of-the-art detection algorithms on realistic channels with spatial correlation. We designed \sys as an iterative algorithm and showed that a carefully chosen degree of flexibility in the model, in addition to leveraging the channel's spectral and temporal correlation, can enable online training at a less or equal computation complexity than other deep-learning based schemes like OAMPNet. \sys is 4--8dB better overall than the classic MMSE detector and it requires 2.5dB lower SNR at the same SER, relative to the second-best detection scheme, OAMPNet, at 10--15$\times$ less computational complexity. Many extensions of \sys  are possible to support, for example, a varying number of transmitters with possibly different modulation schemes. 

 From a hardware perspective, implementation of \sys has its own challenges and requires an in-depth study. For example, the sequential online training algorithm introduced in this paper incurs significant latency which may be traded off with parallel training of multiple sub-carriers at the cost of more training iterations and hence increased complexity. The optimal trade-off depends on the channel coherence time. 
 
We would also like to point out, that assumptions like the \gls{iid} Gaussian property of the channel matrices can  lead to misleading conclusions for MIMO detection performance. Future work should therefore be based on realistic channel models, from either simulation, ray-tracing, or measurements. We release the simulated 3GPP MIMO channels dataset used in this work as an effort in this direction with a hope for more practical benchmarks from the community.

    \bibliography{mimo_refs}

\end{document}